\documentclass
[prc,twocolumn,superscriptaddress,showpacs,twoside,floatfix]{revtex4}
\usepackage{graphicx}
\usepackage{amsmath}

\begin{document}
\title{Two-proton radioactivity and three-body decay. III. Integral formulae for
decay widths in a simplified semianalytical approach.}

\author{L.\ V.\ Grigorenko}
\affiliation{Flerov Laboratory of Nuclear Reactions, JINR, RU-141980 Dubna,
Russia}
\affiliation{Gesellschaft f\"{u}r Schwerionenforschung mbH, Planckstrasse 1,
D-64291, Darmstadt, Germany}
\affiliation{RRC ``The Kurchatov Institute'', Kurchatov sq.\ 1, 123182 Moscow,
Russia}

\author{M.\ V.\ Zhukov}
\affiliation{Fundamental Physics, Chalmers University of Technology,
S-41296 G\"{o}teborg, Sweden}


\begin{abstract}
Three-body decays of resonant states are studied using integral
formulae for decay widths. Theoretical approach with a simplified
Hamiltonian allows semianalytical treatment of the problem. The
model is applied to decays of the first excited $3/2^{-}$ state of
$^{17}$Ne and the $3/2^{-}$ ground state of $^{45}$Fe. The
convergence of three-body hyperspherical model calculations to the
exact result for widths and energy distributions are studied. The
theoretical results for $^{17}$Ne and $^{45}$Fe decays are updated
and uncertainties of the derived values are discussed in detail.
Correlations for the decay of $^{17}$Ne $3/2^-$ state are also
studied.
\end{abstract}


\maketitle


\section{Introduction}

\begin{figure}[ptb]
\includegraphics[width=0.48\textwidth]{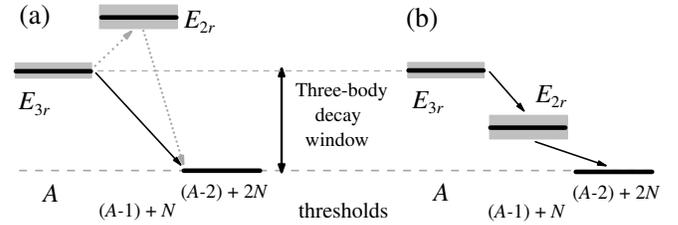}
\caption{Energy conditions for different modes of the two-nucleon emission
(three-body decay): true three-body decay (a), sequential decay (b).}
\label{fig:levels}
\end{figure}

The idea of the ``true'' two-proton radioactivity was proposed about
50 years ago in a classical paper of Goldansky \cite{gol60}. The
word ``true'' denotes here that we are dealing not with a relatively
simple emission of two protons, which becomes possible in every
nucleus above two-proton decay threshold, but with a specific
situation where one-proton emission is energetically (due to the
proton separation energy in the daughter system) or dynamically (due
to various reasons) prohibited. Only simultaneous emission of two
protons is possible in that case (see Fig.\ \ref{fig:levels}, more
details on the modes of the three-body decays can be found in Ref.\
\cite{gri01a}). The dynamics of such decays can not be reduced to a
sequence of two-body decays and from theoretical point of view we
have to deal with a three-body Coulomb problem in the continuum,
which is known to be very complicated.

Progress in this field was quite slow. Only recently a consistent quantum
mechanical theory of the process was developed \cite{gri01a,gri00b,gri03c}, 
which allows to
study the two-proton (three-body) decay phenomenon in a three-body cluster
model. It has been applied to a range of a light nuclear systems
($^{12}$O, $^{16}$Ne \cite{gri02}, $^{6}$Be, $^{8}$Li$^{*}$, $^{9}$Be$^{*}$
\cite{gri02a}, $^{17}$Ne$^{*}$, $^{19}$Mg \cite{gri03}). Systematic exploratory
studies of heavier prospective $2p$ emitters $^{30}$Ar, $^{34}$Ca, $^{45}$Fe,
$^{48}$Ni, $^{54}$Zn, $^{58}$Ge, $^{62}$Se, and $^{66}$Kr \cite{gri03a,gri03c})
have been performed providing predictions of lifetime ranges and possible
correlations among fragments.

Experimental studies of the two-proton radioactivity is presently an
actively developing field. Since the first experimental
identification of $2p$ radioactivity in $^{45}$Fe \cite{pfu02,gio02}
it was also found in $^{54}$Zn \cite{bla05}. Some fingerprints of
the $^{48}$Ni $2p$ decay were observed and the $^{45}$Fe lifetime
and decay energy were measured with improved accuracy \cite{dos05}.
There was an intriguing discovery of the extreme enhancement of the
$2p$ decay mode for the high-spin $21^{+}$ isomer of $^{94}$Ag,
interpreted so far only in terms of the hyperdeformation of this
state \cite{muk06}. New experiments, aimed at more detailed $2p$
decay studies (e.g.\ observation of correlations), are under way at
GSI ($^{19}$Mg), MSU ($^{45}$Fe), GANIL ($^{45}$Fe), and
Jyv\"askyl\"a ($^{94}$Ag).

Several other theoretical approaches were applied to the problem in the recent
years. We should mention the ``diproton'' model \cite{bro91,naz96}, ``R-matrix''
approach \cite{bar01,bar02,bar03,bro03}, continuum shell model \cite{rot06}, and
adiabatic hyperspherical approach of \cite{gar04}. Some issues of a
compatibility between different approaches will be addressed in this work.

Another, possibly very important, field of application of the
two-proton decay studies was shown in Refs.\ \cite{gri05a,gri06}. It
was demonstrated in \cite{gri05a} that the importance of direct
resonant two-proton radiative capture processes was underestimated
in earlier treatment of the rp-process waiting points \cite{gor95}.
The scale of modification of the astrophysical $2p$ capture rates
can be as large as several orders of magnitude in certain
temperature ranges. In paper \cite{gri06} it has been found that
nonresonant E1 contributions to three-body (two-proton) capture
rates can also be much larger than was expected before. The updated
$2p$ astrophysical capture rate for the
$^{15}$O($2p$,$\gamma$)$^{17}$Ne reaction appears to be competing
with the standard $^{15}$O($\alpha$,$\gamma$)$^{19}$Ne breakout
reaction for the hot CNO cycle. The improvements of the $2p$ capture
rates obtained in \cite{gri05a,gri06} are connected to consistent
quantum mechanical treatment of the three-body Coulomb continuum in
contrast to the essentially quasiclassical approach typically used
in astrophysical calculations of three-body capture reactions (e.g.\
\cite{nom85,gor95}).

The growing quality of the experimental studies of the $2p$ decays
and the high precision required for certain astrophysical
calculations inspired us to revisit the issues connected with
different uncertainties and technical difficulties of our studies.
In this work we make the following.
(i) Extend the two-body formalism of the integral formulae for width
to the three-body case. We perform the relevant derivations for the
two-body case to make the relevant approximations and assumptions
explicit.
(ii) Formulate a simplified three-body model which has many dynamical
features similar to the realistic case, but allows the exact
semianalytical treatment and thus makes possible a precise calibration of
three-body calculations. It is also possible to study in great
detail several important dependencies of three-body widths in the
frame of this model.
(iii) Perform practical studies of some systems of interest and
demonstrate a connection between the simplified semianalytical
formalism and the realistic three-body calculations.

The unit system $\hbar=c=1$ is used in the article.


\section{Integral formula for width}


Integral formalisms of width calculations for narrow two-body states
are known for a long time, e.g.\ \cite{har68,kad71}. The prime
objective of those studies was $\alpha$-decay widths. An interesting
overview of this field can be found in the book \cite{kad-book}.
This approach, to our opinion, did not produce novel results as the
inherent uncertainties of the method are essentially the same as
those of the R-matrix phenomenology, which is technically much
simpler (see e.g.\ a discussion in \cite{kad83}). An important nontrivial
application of the integral formalism was calculation of widths for
proton emission off deformed states \cite{bug89,dav98}. There were
attempts to extend the integral formalism to the three-body decays,
using a formal generalization for the hyperspherical space
\cite{dan93,gri01a}. These were shown to be difficult with respect
to technical realisation and to be inferior to other methods
developed in \cite{gri00b,gri01a}.

Here we develop an integral formalism for the three-body
(two-proton) decay width in a different way. However, first we
review the standard formalism to define (clearer) the approximations used.


\subsection{Width definition, complex energy WF}


For decay studies we consider the wave function (WF) with complex pole energy
\[
\tilde{E}_{r}=\tilde{k}_{r}^{2}/(2M)=E_{r}-i \Gamma /2 \quad, \qquad \tilde{k}
_{r}\approx k_{r}-i \Gamma/(2v_{r})\;,
\]
where $v=\sqrt{2E/M}$. The pole solution for Hamiltonian
\[
(H-\tilde{E}_{r})\Psi_{lm}^{(+)}(\mathbf{r})=(T+V-\tilde{E}_{r})\Psi
_{lm}^{(+)}(\mathbf{r})=0
\]
provides the WF with outgoing asymptotic
\begin{equation}
\Psi_{lm}^{(+)}(\mathbf{r})=r^{-1}\psi_{l}^{(+)}(kr)\,Y_{lm}(\hat{r})\;.
\label{wf-pole}
\end{equation}
For single channel two-body problem the pole solution is formed only
for one selected value of angular momentum $l$. In the asymptotic
region
\begin{equation}
\psi_{l}^{(+)}(\tilde{k}_{r}r) \overset{r>R}{=}H_{l}^{(+)}(\tilde{k}
_{r}r)=G_{l}(\tilde{k}_{r}r)+iF_{l}(\tilde{k}_{r}r)\;.
\label{psi-2-plus-ass}
\end{equation}
The above asymptotic is growing exponentially
\[
\psi_{l}^{(+)}(\tilde{k}_{r}r) \overset{r>R}{\sim}\exp[+i \tilde{k}_{r}
r] \approx \exp[+ik_{r}r]\exp[+\Gamma r/(2v_{r})]
\]
as a function of the radius at pole energy. This unphysical growth
is connected to the use of time-independent formalism and could be
reliably neglected for typical \emph{radioactivity} time scale as it
has a noticeable effect at very large distances.

Applying Green's procedure to complex energy WF
\[
\Psi^{(+) \dagger} \left[  (H-\tilde{E}_{r})\Psi^{(+)} \right]  -\left[
(H-\tilde{E}_{r})\Psi^{(+)} \right]  ^{\dagger}\Psi^{(+)}=0
\]
we get for the partial components at pole energy $\tilde{E}_{r}$
\[
i \Gamma \psi_{l}^{(+)\ast}\psi_{l}^{(+)}=\frac{1}{2M}\left[  \psi_{l}^{(+)\ast
}\frac{d^{2}\psi_{l}^{(+)}}{dr^{2}}-\frac{d^{2}\psi_{l}^{(+)\ast}}{dr^{2}}
\psi_{l}^{(+)}\right]  \;.
\]
After radial integration from 0 to $R$ (here and below $R$  denotes
the radius sufficiently large that the nuclear interaction
disappears) we obtain
\begin{equation}
\Gamma=\frac{\left.  \left[  \psi_{l}^{(+)\ast} \! \left(  \frac{d}{dr}\psi
_{l}^{(+)}\right)  -\left(  \frac{d}{dr}\psi_{l}^{(+)\ast}\right) \! \psi
_{l}^{(+)}\right]  \right \vert _{r=R}}{2Mi\;\int_{0}^{R} \left \vert \psi
_{l}^{(+)}\right \vert ^{2}dr}=\frac{j_{l}}{N_{l}}\,,
\label{width-1}
\end{equation}
which corresponds to a definition of the width as a decay
probability (reciprocal of the lifetime):
\[
N=N_{0}\exp[-t/\tau]=N_{0}\exp[-\Gamma t]\;.
\]
The width $\Gamma$ is then equal to the outgoing flux $j_{l}$
through the sphere of sufficiently large radius $R$, divided by
number of particles $N_{l}$ inside the sphere.

Using Eq.\ (\ref{psi-2-plus-ass}) the flux in the asymptotic region could
be rewritten for $\tilde{k}_{r}\rightarrow k_{r}$ in terms of a
Wronskian
\begin{eqnarray}
j_{l}  &  = & \frac{1}{2Mi}\left.  \left[  \psi_{l}^{(+)\ast}\left(  \frac{d}
{dr}\psi_{l}^{(+)}\right)  -\left(  \frac{d}{dr}\psi_{l}^{(+)\ast}\right)
\psi_{l}^{(+)}\right]  \right \vert _{r=R} \nonumber \\
&  = & (k_{r}/M)\;W(F_{l}(k_{r}R),G_{l}(k_{r}R))=v_{r}\;,
\label{flux-wronsk}
\end{eqnarray}
where the Wronskian for real energy functions $F_{l},G_{l}$ is
\[
W(F_{l},G_{l}) = G_l F'_l - G'_l F_l \equiv 1\;.
\]
The effect of the complex energy is easy to estimate (actually
without loss of a generality) in a small energy approximation
\begin{equation}
F_{l}(kr) \overset{kr \rightarrow 0}{\sim}C_{l}(kr)^{l+1},\quad G_{l}
(kr) \overset{kr \rightarrow 0}{\sim} \dfrac{(kr)^{-l}}{(2l+1)C_{l}}\;,
\label{coul-fun-ass}
\end{equation}
where $C_{l}$ is a Coulomb coefficient (defined e.g.\ in Ref.\
\cite{as}). The flux is then
\begin{eqnarray}
j_{l}  &  = & \frac{\tilde{k}_{r}H_{l}^{(-)}(\tilde{k}_{r}^{\ast}r)H_{l}
^{(+)\prime}(\tilde{k}_{r}r)-\tilde{k}_{r}^{\ast}H_{l}^{(-)\prime}(\tilde
{k}_{r}^{\ast}r)H_{l}^{(+)}(\tilde{k}_{r}r)}{2iM}  \nonumber \\
&  = & v_{r} \left(  1-\frac{2l(l+1)}{k_{r}^{2}}\left(  \frac{\Gamma}{2v_{r}
}\right)  ^{2}+l\times o[\Gamma^{3}]\right)  \;. \nonumber
\end{eqnarray}
So, the equality (\ref{flux-wronsk}) is always valid for $l=0$ and for
$l\neq0$ we get
\[
\Gamma\ll\left(  \frac{8}{l(l+1)}\right)  ^{1/2}\,E_{r} \;.
\]


\subsection{Two-body case, real energy WF}


Now we need a WF as real energy $E=k^{2}/2M$ solution of Schr\"odinger equation
\begin{eqnarray}
(H-E)\Psi_{\mathbf{k}}(\mathbf{r})=(T+V^{\text{nuc}} +V^{\text{coul}}-E)
\Psi_{\mathbf{k}} (\mathbf{r})=0
\nonumber \;, \\
\Psi_{\mathbf{k}}(\mathbf{r})=4 \pi \sum \nolimits_{l}i^{l}(kr)^{-1}\psi
_{l}(kr) \sum \nolimits_{m}Y_{lm}^{\ast}(\hat{k})Y_{lm}(\hat{r}) \;,\nonumber
\end{eqnarray}
in S-matrix representation, which means that for $r>R$
\[
\psi_{l}(kr)=\frac{i}{2}\left[  (G_{l}(kr)-iF_{l}(kr))-S_{l}(G_{l}
(kr)+iF_{l}(kr))\right] \;.
\]
At resonance energy $E_{r}$
\[
S_{l}(E_{r})=e^{2i \delta_{l}(E_{r})}=e^{2i\pi/2}=-1
\]
and in asymptotic region, defined by the maximal size of nuclear interaction
$R$,
\[
\psi_{l}(k_{r}r) \overset{r>R}{=}i\,G_{l}(k_{r}r) \;.
\]
At resonance energy we can define a ``quasibound'' WF
$\tilde{\psi}_{l}$ as matching the irregular solution $G_{l}$ and
normalized to unity for the integration in the internal region
limited by radius $R$:
\begin{equation}
\tilde{\psi}_{l}(k_{r}r) =\frac{(-i) \, \psi_{l}(k_{r}r)}{
\left(\int_{0}^{R} \left \vert \psi_{l}(k_{r}x) \right \vert ^{2}dx
\right)^{1/2}} =-i \, \frac{\psi_{l}(k_{r}r)}{{N_l}^{1/2}}.
\label{psi2-tilda}
\end{equation}

Now we introduce an auxiliary Hamiltonian $\bar{H}$ with different short range
nuclear interaction $\bar{V}^{\text{nuc}}$,
\[
(\bar{H}-E)\Phi_{\mathbf{k}}(\mathbf{r})=(T+\bar{V}^{\text{nuc}}
+V^{\text{coul}}-E)\Phi_{\mathbf{k}
}(\mathbf{r})=0 \;,
\]
and also construct other WF in S-matrix representation
\begin{eqnarray}
\Phi_{\mathbf{k}}(\mathbf{r}) & = & 4 \pi \sum \nolimits_{l} i^{l}(kr)^{-1}
\varphi_{l}(kr) \sum \nolimits_{m}Y_{lm}^{\ast}(\hat{k})Y_{lm}(\hat{r})\;,
\nonumber \\
\varphi_{l}(kr) & = & \frac{i}{2}\left[  (G_{l}(kr)-iF_{l}(kr))-\bar{S}_{l}
(G_{l}(kr)+iF_{l}(kr))\right] \;, \nonumber
\end{eqnarray}
for $r>R$. Or in equivalent form:
\begin{equation}
\varphi_{l}(kr)=\exp(i\bar{\delta}_{l})\,\,\left[  F_{l}(kr)\cos(\bar{\delta
}_{l})+G_{l}(kr)\sin(\bar{\delta}_{l})\right]  \,.
\label{auxil-wf-def}
\end{equation}
The Hamiltonian $\bar{H}$ should provide the WF $\Phi_{\mathbf{k}}(\mathbf{r})$
which at energy $E_{r}$ is sufficiently far from being a resonance WF and for
this WF $\bar{\delta}_{l}(E_r)\sim0 $.

For real energy WFs $\Psi_{\mathbf{k}}(\mathbf{r})$ and $\Phi_{\mathbf{k}}
(\mathbf{r})$ we can write:
\begin{eqnarray}
\Phi_{\mathbf{k}}(\mathbf{r})^{\dagger}\left[  (H-E)\Psi_{\mathbf{k}}
(\mathbf{r})\right]  -\left[ (\bar{H}-E) \Phi_{\mathbf{k}}(\mathbf{r})\right]
^{\dagger}\Psi_{\mathbf{k}}(\mathbf{r})=0 \,, \nonumber \\
\varphi_{l}^{\ast}(V-\bar{V})\psi_{l}=\frac{1}{2M}\left[  \varphi_{l}^{\ast
}\left(  \frac{d^{2}}{dr^{2}}\psi_{l}\right)  -\left(  \frac{d^{2}}{dr^{2}
}\varphi_{l}^{\ast}\right)  \psi_{l}\right]\,,
\end{eqnarray}
For WFs taken at resonance energy $E_{r}$ this expression provides
\begin{eqnarray}
2M \int_{0}^{R}\varphi_{l}^{\ast}(V-\bar{V})\psi_{l}dr =
2Mi{N_{l}}^{1/2} \int_{0}
^{R}\varphi_{l}^{\ast}(V-\bar{V})\tilde{\psi}_{l}dr \nonumber \\
=\exp(-i \bar{\delta}_{l})\cos(\bar{\delta}_{l})  \;k_{r}\; W(F_{l}
(k_{r}R),G_{l}(k_{r}R))  \;,
\label{int-via-wronsk} \\
{N_{l}}^{1/2}=\frac{-i \exp(-i \bar{\delta}_{l}) \cos(\bar{\delta}_{l}) \,
k_{r}} {2M \int_{0}^{R}\varphi_{l}^{\ast}(V-\bar{V})\tilde{\psi}_{l}dr} \;.
\nonumber
\end{eqnarray}
From Eqs.\ (\ref{width-1}), (\ref{flux-wronsk}), (\ref{psi2-tilda})
and the approximation $\psi _{l}^{(+)}\approx\psi_{l}$ it follows
that
\begin{eqnarray}
\Gamma=\frac{v_{r}}{\int_{0}^{R} \left \vert \psi_{l}^{(+)} \right
\vert ^{2} dr} \approx \frac{v_{r}}{\int_{0}^{R} \left \vert
\psi_{l} \right \vert ^{2} dr}=\frac{v_{r}}{\left \vert
N_{l}^{1/2}\right \vert^{2}} \;,
\nonumber \\
\Gamma=\frac{4}{v_{r}\cos^{2}(\bar{\delta}_{l})} \left \vert \int_{0}^{R}
\varphi_{l}^{\ast}(V-\bar{V})\tilde{\psi}_{l}dr \right \vert ^{2} \;.
\label{width-2}
\end{eqnarray}

So, the idea of the integral method is to define the internal normalizations for
the WF with resonant boundary conditions (this is equivalent to determination of
the outgoing flux for normalized ``quasibound'' WF) by the help of the
eigenfunction of the auxiliary Hamiltonian, which has the same long-range
behaviour and differs only in the compact region.


\section{Alternative derivation}


Let us reformulate the derivation of Eq.\ (\ref{width-2}) in a more general way, 
so that the detailed knowledge of the WF structure for $\psi_l$ and 
$\psi^{(+)}_l$ is not required. It would allow a straightforward extension of 
the formalism to the three-body case. We start from Schr\"odinger equation in 
continuum with solution $\Psi^{(+)}$ at the pole energy $\tilde{E}_{r}=E_{r}+i\, 
\Gamma/2$:
\begin{equation}
\left(  H-\tilde{E}_{r}\right)  \Psi^{(+)}=\left(  T+V-\tilde{E}_{r}\right)
\Psi^{(+)}=0 \;.
\label{alt-der-0}
\end{equation}
Then we rewrite it identically via the auxiliary Hamiltonian
$\bar{H}=T+\bar{V}$
\begin{eqnarray}
\left(  H+\bar{V}-V-\tilde{E}_{r}\right)  \Psi^{(+)}=\left(  \bar{V}-V \right)
\Psi^{(+)}\nonumber \\
\left(  \bar{H}-E_{r}\right)  \Psi^{(+)}=\left(  \bar{V}-V+i \, \Gamma/2 \right)
\Psi^{(+)}.
\label{alt-der-1}
\end{eqnarray}
Thus we can use the real-energy Green's function $\bar{G}_{E_{r}}$ of auxiliary
Hamiltonian $\bar{H}$ to ``regenerate'' the WF with outgoing asymptotic
\begin{equation}
\bar{\Psi}^{(+)}=\bar{G}_{E_{r}}^{(+)}\left(  \bar{V}-V+i \, \Gamma /2 \right)
\Psi^{(+)} \;.
\label{alt-der-2}
\end{equation}
%
%
At this point in Eq.\ (\ref{alt-der-2})
$\bar{\Psi}^{(+)} \equiv \Psi^{(+)}$ and the bar in the notation for
``corrected'' WF $\bar{\Psi}^{(+)}$ is introduced for later use to
distinguish it from the ``initial'' WF $\Psi^{(+)}$ [the one before
application of Eq.\ (\ref{alt-der-2})]. Further assumptions we
should consider separately in two-body and three-body cases.


\subsection{Two-body case}


To define the width $\Gamma$ by Eq.\ (\ref{width-1}) we need to know
the complex-energy solution $\Psi^{(+)}$ at pole energy. For narrow
states $\Gamma\ll E_{r}$ this solution can be obtained in a
simplified way using the following approximations.

(i) For narrow states we can always choose the auxiliary Hamiltonian in such a
way that $\Gamma\ll\bar{V}-V$, and we can assume $\Gamma \rightarrow 0$ in the
Eq.\ (\ref{alt-der-2}).

(ii) Instead of complex-energy solution $\Psi^{(+)}$ in the
right-hand side of (\ref{alt-der-2}) we can use the normalized
real-energy quasibound solution $\tilde{\Psi}$ defined for one
real resonant value of energy $E_{r}=k_{r}^{2}/2M$
\[
N_{l}=\int d \Omega \int_{0}^{R}dr\;r^{2} \left \vert \tilde{\Psi}_{lm}
(\mathbf{r}) \right \vert ^{2} \equiv 1  \;.
\]

So, the Eq.\ (\ref{alt-der-2}) is used in the form
\begin{equation}
\bar{\Psi}_{lm}^{(+)}=\bar{G}_{E_{r}}^{(+)}\left(  \bar{V}-V \right)
\tilde{\Psi}_{lm} \;.
\label{alt-der-3}
\end{equation}
The solution $\bar{\Psi}^{(+)}$ is matched to function
\begin{equation}
h_l^{(+)}(kr)=G_{l}(kr)+iF_{l}(kr) \;,
\label{eq:wf-out-def}
\end{equation}
while the solution $\tilde{\Psi}$ is matched to function $G_{l}$.
For deep subbarrier energies it is reasonable to expect that in the
internal region $r \leq R$
\[
G_{l} \gg F_{l}\; \rightarrow \; \left \Vert
\operatorname{Re}[\tilde{\Psi}^{(+)}] \right \Vert \approx \left \Vert
\tilde{\Psi}\right \Vert
\gg \left \Vert \operatorname{Im}[\tilde{\Psi}^{(+)}]\right \Vert \;.
\]
In the single channel case it can be shown by direct calculation
that an approximate equality
\[
\frac{MR^{2}\Gamma}{\pi}\left \Vert \operatorname{Re}[\tilde{\Psi}
^{(+)}]\right \Vert \gtrsim \left \Vert \operatorname{Im}[\tilde{\Psi}
^{(+)}]\right \Vert
\]
holds in the internal region and thus for narrow states $\Gamma\ll E_{r}$ the
approximation (\ref{alt-der-2}) $\rightarrow$ (\ref{alt-der-3}) should be very
reliable.

To derive Eq.\ (\ref{width-2}) the WF with outgoing asymptotic is generated
using the Green's function of the auxiliary Hamiltonian $\bar{H}$ and the
``transition potential'' ($V-\bar{V}$). The standard two-body Green's function
is
\begin{eqnarray}
\bar{G}_{k^{2}/(2m)}^{(+)}(\mathbf{r},\mathbf{r}^{\prime}) &  = & \frac{2M}
{krr^{\prime}}\sum \limits_{l}\left\{
\begin{array}
[c]{c}
\varphi_{l}(kr)\,h_{l}^{(+)}(kr^{\prime}),r\leq r^{\prime}\\
h_{l}^{(+)}(kr)\varphi_{l}(kr^{\prime}),r>r^{\prime}
\end{array} \right\}  \nonumber \\
& \times & \sum \nolimits_{m}Y_{lm}(\hat{r})Y_{lm}^{\ast}(\hat{r}^{\prime}) \;,
\label{green-fun-2}
\end{eqnarray}
where the radial WFs $h_{l}^{(+)}$ and $\varphi_{l}$ of the
auxiliary Hamiltonian are defined in (\ref{eq:wf-out-def}) and
(\ref{auxil-wf-def}).
\[
\bar{\Psi}_{lm}^{(+)}(\mathbf{r})=\int d \mathbf{r}^{\prime}\;\bar{G}_{k^{2}
/(2m)}^{(+)}(\mathbf{r},\mathbf{r}^{\prime})\left(  \bar{V}-V \right)
\tilde{\Psi}_{l^{\prime}m^{\prime}}(\mathbf{r}^{\prime}) \;.
\]
For the asymptotic region $r>R$
\begin{eqnarray}
\bar{\Psi}_{lm}^{(+)}(\mathbf{r}) &  = & \frac{2M}{k_{r}r}h_{l}^{(+)}
(k_{r}r)Y_{lm}(\hat{r}) \nonumber \\
&  \times & \int_{0}^{R}dr'\;\varphi_{l}(k_{r}r')\,\left(  \bar{V}-V\right)
\tilde{\psi}_{l}(k_{r},r') \;. \nonumber
\end{eqnarray}
The outgoing flux is then calculated [see Eq.\ (\ref{flux-wronsk})]
\[
j_{l}=\frac{R^{2}}{2l+1}\sum_{m}\int d\Omega\left.  \frac{1}{M}
\operatorname{Im}\left[  \bar{\Psi}_{lm}^{(+)\ast}(\mathbf{r})\nabla
\bar{\Psi}_{lm}^{(+)}\right]  \right\vert _{r=R} \;.
\]
As far as function $\tilde{\Psi}$ is normalized by construction then
\begin{equation}
\Gamma \equiv j_{l}=\frac{4}{v_{r}}\left\vert \int_{0}^{R} dr \; \varphi_{l}
(k_{r}r) \, \left( \bar{V}-V \right) \tilde{\psi}_{l} (k_{r},r) \right \vert
^{2} \;.
\label{width-2-2}
\end{equation}
Note, that this equation differs from Eq.\ (\ref{width-2}) only by a factor
$1/(\cos^{2}[\bar{\delta}_{l}])$ which should be very close to unity for
sufficiently high barriers.


\subsection{Simplified model for three-body case}


\begin{figure}[tb]
\includegraphics[width=0.45\textwidth]{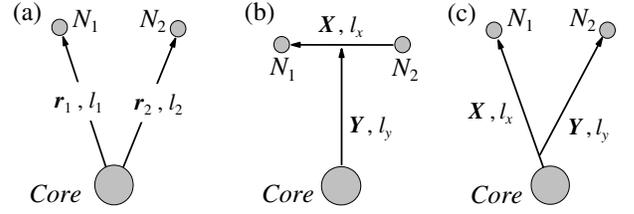}
\caption{Single particle coordinate systems: (a) ``V'' system typical for a
shell model. In the Jacobi ``T'' system (b), ``diproton'' and core are
explicitly in configurations with definite angular momenta $l_{x}$ and $l_{y}$.
For a heavy core the Jacobi ``Y'' system (c) is close to the single particle
system (a).}
\label{fig:jac-sys}
\end{figure}

In papers \cite{gri00b,gri01a} the widths for three-body decays were
defined by the following procedure. We solve numerically the problem
\[
\left(  H-E_{3r} \right)  \tilde{\Psi}=0
\]
with \emph{some} box boundary conditions (e.g.\ zero or quasibound
in diagonal channels at large distances) getting the WF $\tilde{\Psi}$
normalized in the finite domain and the value of the real resonant energy 
$E_{3r}$. Thereupon we search for the outgoing solution $\Psi^{(+)}$ of the 
equation
\[
\left(  H-E_{3r} \right)  \Psi^{(+)}=-i\Gamma/2\;\tilde{\Psi}
\]
with approximate boundary conditions of three-body Coulomb problem (see Ref.\
\cite{gri01a} for details) and arbitrary $\Gamma$. The width is then defined as
the flux through the hypersphere of the large radius divided by normalization
within this radius:
\begin{equation}
\Gamma=\frac{j}{N}=\frac{\int d\Omega_{5}\;\left.  \Psi^{(+)\ast}\rho
^{5/2}\frac{d}{d\rho}\rho^{5/2}\Psi^{(+)}\right\vert _{\rho=\rho_{\max}}
}{M\;\int d \Omega_{5}\int_{0}^{\rho_{\max}}\rho^{5}d\rho\;\left\vert
\Psi^{(+)}\right\vert ^{2}}
\label{wid-3b-def}
\end{equation}
The 3-body WF with outgoing asymptotic is
\begin{equation}
\Psi_{JM}^{(+)}(\rho,\Omega_5)=\rho^{-5/2} \sum \nolimits _{K\gamma}
\chi_{K\gamma}^{(+)}(\rho)\,{\cal J}^{JM}_{K\gamma}(\Omega_5)\;,
\label{wf-3b-out}
\end{equation}
where the definitions of the hyperspherical variables $\rho$, $\Omega_5$ and
hyperspherical harmonics ${\cal J}^{JM}_{K\gamma}$ can be found in Ref.\
\cite{gri03c}.

Here we formulate the simplified three-body model in the way which,
on one hand, keeps the important dynamical features of the
three-body decays (typical sizes of the nuclear potentials, typical
energies in the subsystems, correct ratios of masses, etc.), and, on
the other hand, allows a semianalytical treatment of the problem.
Two types of approximations are made here.

The three-body Coulomb interaction is
\begin{equation}
V^{\text{coul}}=\frac{Z_{1}Z_{2}\alpha}{X} + \frac{Z_{1} Z_{3}\alpha}{\left
\vert
\mathbf{Y}+
\frac{A_2\mathbf{X}}{A_1+A_2}\right \vert} + \frac{Z_{2} Z_{3}\alpha}{\left
\vert \mathbf{Y}-
\frac{A_1\mathbf{X}}{A_1+A_2} \right \vert} \;,
\label{eq:coul-pot-3}
\end{equation}
where $\alpha$ is the fine structure constant. By convention, see e.g.\ Fig.\
\ref{fig:jac-sys}, in the ``T'' Jacobi system the core is particle number 3 and
in ``Y'' system it is particle number 2. We assume that the above potential can
be approximated by Coulomb terms which depend on Jacobi variables $X$ and $Y$
only:
\[
V_{x}^{\text{coul}}(X)=\frac{Z_{x}\alpha}{X}\quad,\quad
V_{y}^{\text{coul}}(Y)=\frac
{Z_{y}\alpha}{Y}\;,
\]
(in reality for the small $X$ and $Y$ values the Coulomb formfactors of the
homogeneously charged sphere with radius $r_{sph}$ are always used).
The effective charges $Z_{x}$ and $Z_{y}$ could be considered in two ways.

\begin{enumerate}
\item We can neglect one of the Coulomb interactions. This approximation is
consistent with physical situation of heavy core and treatment of
two final state interactions. Such a situation presumes that Jacobi
``Y'' system is preferable and there is a symmetry in the treatment
of the $X$ and $Y$ coordinates, which are close to shell-model
single particle coordinates.
\begin{equation}
Z_{x}=Z_{1}Z_{\text{core}}\quad,\qquad Z_{y}=Z_{2}Z_{\text{core}}\;.
\label{app-coul-1}
\end{equation}
Further we refer this approximation as ``no $p$-$p$ Coulomb'' case,
as typically the proton-proton Coulomb interaction is neglected
compared to Coulomb interaction of a proton with heavy core.

\item We can also consider two particles on the $X$ coordinate as one
single particle. The Coulomb interaction in $p$-$p$ channel is thus
somehow taken into account effectively via a modification of the
$Z_{y}$ charge:
\begin{equation}
Z_{x}=Z_{1}Z_{\text{core}}\;,\qquad Z_{y}=Z_{2}(Z_{\text{core}}+Z_{1}) \;.
\label{app-coul-2}
\end{equation}
Below we call this situation as ``effective $p$-$p$ Coulomb'' case.

\end{enumerate}

For nuclear interactions we can assume that

\begin{enumerate}

\item There is only one nuclear pairwise interaction and
\begin{eqnarray}
H   =T+V_{3}(\rho)+V_{x}^{\text{coul}}(X) \nonumber \\
  +V_{x}^{\text{nuc}}(X)+V_{y}^{\text{coul}}(Y)\;,\nonumber \\
\Delta V(X,Y)=V_{y}^{\text{nuc}}(Y)-V_{3}(\rho)\;.
\label{app-nucl-1}
\end{eqnarray}
This approximation is good for methodological purposes as it allows
to focus on one degree of freedom and isolate it from the others.
From physical point of view it could be reasonable if only one FSI
is strong \footnote{A realistic example of this situation is the
case of ``E1'' (coupled to the ground state by the E1 operator)
continuum considered in Ref.\ \cite{gri06}. This case is relevant to
the low energy radiative capture reactions, important for
astrophysics, but deal with nonresonant continuum only.}, or we have
reasons to think that decay mechanism associated with this
particular FSI is dominating. Potential $V_{y}^{\text{nuc}}(Y)$ in
the auxiliary Hamiltonian (\ref{aux-ham-3}) is ``unphysical'' in
that case and can be put zero \footnote{Interesting numerical
stability test is a variation of the ``unphysical'' (for OFSI
approximation) potential $V_{y}^{\text{nuc}}(Y)$ in the auxiliary
Hamiltonian (\ref{aux-ham-3}). It can be used for numerical tests of
the procedure as it should not influence the width. Really, for
variation of this potential from weak attraction (we should not
allow an unphysical resonance into decay window) to strong repulsion
(scale of the variation is tens of MeV for potential with some
typical radius) the width is varied only within couple of percents.
This shows high numerical stability of the procedure.}. We further
refer this model as ``one final state interaction'' (OFSI).

\item We can consider two final state interactions (TFSI). Simple form of the
Green's function in that case can be preserved only if the core mass
is considered as infinite (the $X$ and $Y$ coordinates in the Jacobi
``Y'' system coincide with single-particle core-$p$ coordinates). In
that case both pairwise interactions $V_{x}^{\text{nuc}}(X)$ and
$V_{y}^{\text{nuc}}(Y)$ are treated as ``physical'', that means that they
are both present in the initial and in the auxiliary Hamiltonians. Thus only
three-body potential ``survive'' the $\bar{V}-V$ subtraction:
\begin{eqnarray}
H   =T+V_{3}(\rho)+V_{x}^{\text{coul}}(X)+V_{x}^{\text{nuc}}(X)  \nonumber \\
 +V_{y}^{\text{coul}}(Y)+V_{y}^{\text{nuc}}(Y)\;,\nonumber \\
\Delta V(X,Y)=-V_{3}(\rho)\;.
\label{app-nucl-2}
\end{eqnarray}

\end{enumerate}

The three-body potential is used in this work in Woods-Saxon form
\begin{equation}
V_{3}(\rho)=V_{3}^{0}\left(1+\exp\left[ (\rho-\rho_{0})/a_{\rho} \right]
\right)^{-1}\;,
\label{pot-3b}
\end{equation}
with $\rho_{0}=5$ fm for $^{17}$Ne, $\rho_{0}=6$ fm for $^{45}$Fe
\footnote{These values can be evaluated as typical nuclear radius
for the system multiplied by $\sqrt{2}$: $3.53 \sqrt{2} \approx 5$
and $4.29 \sqrt{2} \approx 6$.}, and a small value of diffuseness
parameter $a_{\rho}=0.4$ fm. Use of such three-body potential is an
important difference from our previous calculations, where it was
utilized in the form
\begin{equation}
V_{3}(\rho)=V_{3}^{0} \left(1+(\rho/\rho_{0})^{3}\right)^{-1} \;,
\label{pot-3b-old}
\end{equation}
which provides the long-range behaviour $\sim \rho^{-3}$. Such an asymptotic in
$\rho$ variable is produced by short-range pairwise nuclear interactions and
thus the interpretation of three-body potential (\ref{pot-3b-old}) is
phenomenological taking into account those components of pairwise interactions
which were omitted for some reasons in calculations. In this work the aim of the
potential $V_{3}$ is different. On one hand we would like to keep the three-body
energy fixed while the properties (and number) of pairwise interactions are
varied. On the other hand we do not want to change the properties of the Coulomb
barriers beyond the typical nuclear distance (this is achieved by the small
diffuseness of the potential). Thus this potential is phenomenological taking
into account interactions that act only when both valence nucleons are close to
the core (both move in the mean field of the nucleus).

The auxiliary Hamiltonian is taken in the form that allows a separate treatment
of $X$ and $Y$ variables
\begin{equation}
\bar{H}=T+V_{x}^{\text{coul}}(X)+V_{x}^{\text{nuc}}(X)+V_{y}^{\text{coul}}(Y)+V_
{y}^{\text{nuc}}
(Y)
\label{aux-ham-3}
\end{equation}
In this formulation of the model the Coulomb potentials are fixed as shown
above. The nuclear potential $V_{x}^{\text{nuc}}(X)$ [$V_{y}^{\text{nuc}}(Y)$ if
present] defines the position of the state in the $X$ [$Y$] subsystem. The
three-body potential $V_{3}(\rho)$ defines the position of the three-body state,
which is found using the three-body HH approach of \cite{gri01a,gri03c}. After
that a new WF with outgoing asymptotic is generated by means of the three-body
Green's function which can be written for (\ref{aux-ham-3}) in a factorized form
(without paying attention to the angular coupling)
\[
G_{E_{3r}}^{(+)}(\mathbf{XY,X}^{\prime}\mathbf{Y}^{\prime})=\frac{1}{2\pi i}
\int_{-\infty}^{\infty} \!\!\! d E_x \,G_{E_x}^{(+)}(\mathbf{X,X}
^{\prime})\,G_{E_y}^{(+)}(\mathbf{Y,Y}^{\prime}) ,
\]
where $E_{3r}=E_x+E_y$ ($E_x$, $E_x$ are energies of subsystems).
The two-body Green's functions in the expressions above are defined
as in (\ref{green-fun-2}) via eigenfunctions of the subhamiltonians
\[
\left\{
\begin{array}
[c]{l}
\bar{H}_{x}-E_x =T_{x}+V_{x}^{\text{coul}}(X)+V_{x}^{\text{nuc}}(X)-E_x \\
\bar{H}_{y}-E_y=T_{y}+V_{y}^{\text{coul}}(Y)+V_{y}^{\text{nuc}} (Y)-E_y
\end{array}
\right. \;.
\]
In the OFSI case the nuclear potential in the ``Y'' subsystem should be put
$V_{y}^{\text{nuc}}(Y) \equiv 0$. The ``corrected'' continuum WF
$\bar{\Psi}^{(+)}$ is
\begin{eqnarray}
\bar{\Psi}^{(+)}(\mathbf{X},\mathbf{Y}) & = & \frac{1}{2\pi i}\int d
\mathbf{X}^{\prime} d
\mathbf{Y}^{\prime}\int \nolimits_{- \infty }^{ \infty }d E_x G_{E_x}^{(+)}
(\mathbf{X}, \mathbf{X}^{\prime})  \nonumber \\
&  \times & G_{E_y}^{(+)}(\mathbf{Y},\mathbf{Y}^{\prime})\;\Delta V(X',Y')\;
\Psi^{(+)} (\mathbf{X'Y'})  \nonumber
\end{eqnarray}
The ``initial'' solution $\Psi^{(+)}$ of Eq.\ (\ref{wf-3b-out}) rewritten in the
coordinates $X$ and $Y$ is
\begin{equation}
\Psi^{(+)}_{JM}(\mathbf{X},\mathbf{Y})=\frac{ \varphi _{Ll_{x}l_{y}S}(X,Y)}
{XY}\left[  \left[  l_{y}\otimes l_{x}\right]  _{L}\otimes S \right] _{JM}
\label{gs-wf}
\end{equation}
%

\begin{figure}[ptb]
\includegraphics[width=0.47\textwidth]{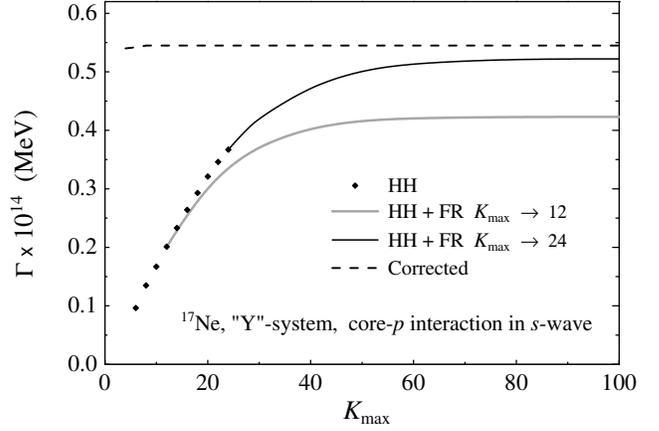}
\caption{Convergence of the $^{17}$Ne width in a simplified model in the ``Y''
Jacobi system. One final state interaction model with experimental position
$E_{2r}=0.535$ KeV of the $s$-wave two-body resonance. Diamonds show the results
of dynamic HH calculations. Solid curves correspond to calculations with
effective FR potentials.}
\label{fig:gam-s}
\end{figure}

The asymptotic form of the "corrected" continuum WF $\bar{\Psi}^{(+)}_{JM}$ is
\begin{eqnarray}
\bar{\Psi}^{(+)}_{JM}(\mathbf{X,Y}) =  \frac{1}{2 \pi i} \, \frac{E_{3r}}{XY}
\int \nolimits_{0} ^{1}d \varepsilon \;
\frac{4}{v_{x}(\varepsilon)v_{y}(\varepsilon)} \;A(\varepsilon)
  \nonumber  \\
 \times e^{ik_{x}(\varepsilon)X+ik_{y}(\varepsilon)Y}\;  \left[\left[
l_{y}\otimes l_{x}\right]  _{L}\otimes S \right] _{JM}
\nonumber  \\
E_x=\varepsilon E_{3r}  \; ;\quad E_y=(1-\varepsilon) E_{3r}  \; ;\quad
v_{i}(\varepsilon)  =  \sqrt{2 E_i /M_{i}} \nonumber  \\
A(\varepsilon)  =  \int \nolimits_{0}^{R}dX^{\prime} \int \nolimits_{0}^{R}
dY^{\prime} \; \varphi_{l_{x}}(k_{x}(\varepsilon)X^{\prime}
)\;\varphi_{l_{y}}(k_{y}(\varepsilon)Y^{\prime}) \nonumber  \\
  \times  \Delta V(X^{\prime},Y^{\prime})\;\varphi_{Ll_{x}l_{y}S}(X^{\prime
},Y^{\prime})\,.  \;
\label{coef-a}
\end{eqnarray}
The ``corrected'' outgoing flux $j_c$ can be calculated on the
sphere of the large radius for any of two Jacobi variables. E.g.\
for $X$ coordinate we have \footnote{The derivation of the flux here
is given in a schematic form. The complete proof is quite bulky to
be provided in the limited space. We would mention only that it is
easy to check directly that the derived expression for flux
preserves the continuum normalization.}
\begin{eqnarray}
j_{c}(E_{3r}) & = & \left.  \operatorname{Im} \left[  X^{2} \! \!
\int d \Omega_{x} \! \! \int d \mathbf{Y}\left( \bar{\Psi}^{(+)\ast}
\frac{\nabla_{X}}{M_{x}} \bar{\Psi}^{(+)}\right) \right]  \right
\vert _{X \rightarrow \infty} \nonumber
\\
& = & E_{3r}^2 \int \nolimits_{0}^{1}d \varepsilon \;\frac{A^{\ast}
(\varepsilon)} {2\pi} \frac{4}{v_{x}v_{y}} \int \nolimits_{0}^{1} d
\varepsilon^{\prime}\;\frac{k_{x}(\varepsilon)}
{M_{x}}\frac{A(\varepsilon^{\prime})} {2\pi}  \nonumber \\
&  \times & \frac{4} {v_{x}^{\prime}v_{y}^{\prime}}\;2\pi \, \delta(k_{y}
(\varepsilon^{\prime}) -k_{y}(\varepsilon)) \; .
\label{wid-3b-def-c}
\end{eqnarray}
Values $v'_i$ above denote $v_i(\varepsilon')$. The flux is obtained as
\begin{equation}
j_{c}(E_{3r})=\frac{8}{\pi} \, E_{3r} \int \nolimits_{0}^{1} d \varepsilon \;
\frac{1}{v_{x}(\varepsilon)v_{y}(\varepsilon)} \left \vert A(\varepsilon) \right
\vert ^{2} \;.
\label{corr-flux}
\end{equation}
In principle as we have seen above that the widths obtained with both fluxes
Eqs.\ (\ref{wid-3b-def}) and (\ref{corr-flux})  should be equal
\begin{equation}
\Gamma=\frac{j}{N} \equiv \Gamma_{c}=\frac{j_{c}}{N}\;.
\label{corr-width}
\end{equation}
This is the idea of calibration procedure for the simplified three-body model.
The convergence of the HH method (for WF $\Psi^{(+)}_{JM}$) is \emph{expected}
to be fast in the internal region and much slower in the distant subbarrier
region. This should be true for the width $\Gamma$ calculated in the HH method.
However, the procedure for calculation of the ``corrected'' width $\Gamma_{c}$
is exact under the barrier and it is sensitive only to HH convergence in the
internal region, which is achieved easily. Below we demonstrate this in
particular calculations.


\section{Decays of the $^{17}$Ne $3/2^{-}$ and $^{45}$Fe $3/2^{-}$ states in a
simplified model}
\label{sec:dec-simp}


In this Section when we refer widths of $^{17}$Ne and $^{45}$Fe we always mean
the $^{17}$Ne $3/2^{-}$ state ($E_{3r}=0.344$ MeV) and the $^{45}$Fe $3/2^{-}$
ground state ($E_{3r}=1.154$ MeV) calculated in a very simple models. We expect
that important regularities found for these models should be true also in
realistic calculations. However, particular values obtained in realistic models
may differ significantly, and this issue is considered specially in the Section
\ref{sec:three-b}.

To keep only the most significant features of the systems we assume
pure $sd$ structure $(l_{x}=0,l_{y}=2)$ for $^{17}$Ne and pure $p^2$
structure $(l_{x}=1,l_{y}=1)$ for $^{45}$Fe in "Y" Jacobi system
(see Fig. \ref{fig:jac-sys}). Spin dependencies of the interactions
are neglected. The Gaussian formfactor
\[
V_{i}^{\text{nuc}}(r)=V_{i0}\exp[-(r/r_{0})^{2}]\;,
\]
where $i=\{x,y\}$, is taken for $^{17}$Ne (see Table
\ref{tab:poten-ne}), and a standard Woods-Saxon formfactor is used
for $^{45}$Fe (see Table \ref{tab:poten-fe}),
\begin{equation}
V_{i}^{\text{nuc}}(r)=V_{i0}\left[  1+\exp[(r-r_{0})/a]\right]  ^{-1}\;.
\label{woods-saxon}
\end{equation}

\begin{figure}[ptb]
\includegraphics[width=0.47\textwidth]{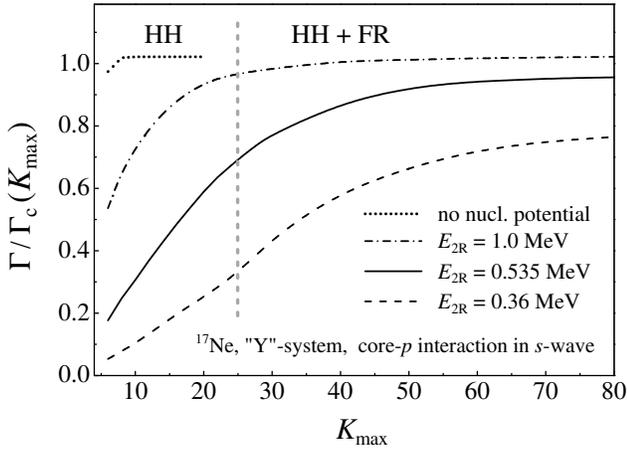}
\caption{Convergence of widths in OFSI model for different positions
$E_{2r}$ of the two-body resonance in the core-$p$ channel (Jacobi
``Y'' system). For $K_{\max}>24$ the value of $K_{\max}$ denote the
size of the basis for Feshbach reduction to $K_{\max}=24$.}
\label{fig:gam-ex-conv}
\end{figure}

The simplistic structure models can be expected to overestimate the widths.
There should be a considerable weight of $d^2$ component $(l_{x}=2,l_{y}=2)$ in
$^{17}$Ne and  $f^2$ component $(l_{x}=3,l_{y}=3)$ in $^{45}$Fe. Also the
spin-angular coupling should lead to splitting of the single-particle strength
and corresponding reduction of the width estimates (e.g.\ we assume one $s$-wave
state at 0.535 keV in the ``X'' subsystem of $^{17}$Ne while in reality there
are two $s$-wave states in $^{16}$F: $0^{-}$ at 0.535 MeV and $1^{-}$ at 0.728
MeV). Thus the results of the simplified model should most likely be regarded as
upper limits for widths.

\begin{figure}[ptb]
\includegraphics[width=0.47\textwidth]{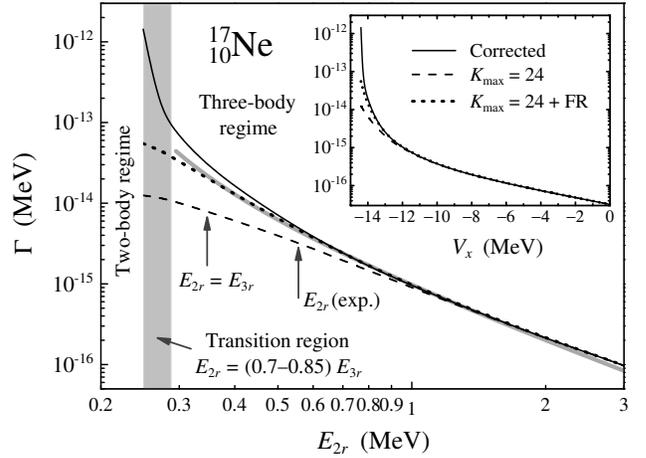}
\caption{Width of the $^{17}$Ne $3/2^-$ state as a function of
two-body resonance position $E_{2r}$. Dashed, dotted and solid lines
show cases of pure HH calculations with $K_{\max}=24$, the same but
with Feshbach reduction from $K_{\max}=100$, and the corrected width
$\Gamma_{c}$. Inset shows the same, but as a function of the
potential depth parameter $V_{x0}$. Gray area shows the transition
region from three-body to two-body decay regime. The gray curve
shows simple analytical dependence of Eq.\ (\ref{eq:anal-e3r-e2r}).}
\label{fig:gam-ot-ex-vx}
\end{figure}


\subsection{One final state interaction --- core-$p$ channel}
\label{sec:ofsi}


\begin{table}[b]
\caption{Parameters for $^{17}$Ne calculations. Potential parameters
for $^{15}$O+$p$ channel in $s$-wave ($V_{x0}$ in MeV, $r_{0}=3.53$
fm) and $^{16}$F+$p$ channel in $d$-wave ($V_{y0}$ in MeV). Radius
of the charged sphere is $r_{sph}=3.904$ fm. Widths $\Gamma_{i}$  of
the state in the subsystem and experimental width values
$\Gamma_{\text{exp}}$ for really existing at these energies states
are given in keV. The corrected three-body width $\Gamma_{c}$ is
given in the units $10^{-14}$ MeV. TFSI calculations with $d$-wave
state at $1.2$ MeV are made with $s$-wave state at $0.728$ MeV.}
\label{tab:poten-ne}
\begin{ruledtabular}
\begin{tabular}[c]{cccccc}
$E_{2r}$ & $l_x$ $(l_y)$ & $V_{x0}$ $(V_{y0})$ & $\Gamma_{x}$ $(\Gamma_{y})$ &
$\Gamma_{\text{exp}}$ & $\Gamma_{c}$  \\
\hline
$0.258$ & 0 & $-14.4$  & $0.221$ &       &  $144$\\
$0.275$ & 0 & $-14.35$ & $0.355$ &       &  $16.6$\\
$0.292$ & 0 & $-14.3$  & $0.544$ &       &  $7.75$\\
$0.360$ & 0 & $-14.1$  & $2.09$  &       &  $2.34$\\
$0.535$ & 0 & $-13.55$ & $17.9$  & 25(5) \cite{ste06} &  $0.545$\\
$0.728$ & 0 & $-12.89$ & $ 72.0$ & 70(5) \cite{ste06} &  $0.211$ \\
$1.0$   & 0 & $-12.0$  & $252$   &       & $0.093$\\
$2.0$   & 0 & $-9.0$   & $\sim 1500$ &   & $0.021$\\
\hline
$0.96$  & 2 & $-87.06$ & $3.5$   & 6(3) \cite{ste06} & $4.73$\footnotemark[1] \\
$1.256$ & 2 & $-85.98$ & $12.2$  & $<15$ \cite{ajz86}& $2.0$\footnotemark[1] \\
$0.96$  & 2 & $-66.46$ & $3.6$   & 6(3) \cite{ste06} & $1.37$\footnotemark[2] \\
$1.256$ & 2 & $-65.4$  & $13.7$  & $<15$ \cite{ajz86}& $0.584$\footnotemark[2]
\end{tabular}
\end{ruledtabular}
\footnotetext[1]{This is TFSI calculation with ``no $p$-$p$'' Coulomb,
$r_{0}=2.75$ fm.}
\footnotetext[2]{This is TFSI calculation with ``effective'' Coulomb,
$r_{0}=3.2$ fm.}
\end{table}

First we take into account only the $0.535$ MeV $s$-wave two-body
resonance in the $^{16}$F subsystem (this is the experimental energy
of the first state in $^{16}$F). Convergence of the $^{17}$Ne width
in a simplified model for Jacobi ``Y'' system is shown in Fig.\
\ref{fig:gam-s}. The convergence of the corrected width $\Gamma_{c}$
as a function of $K_{\max}$ is very fast: $K_{\max}>8$ for the width
is stable within $\sim 1\%$. For maximal achieved in the fully
dynamic calculation $K_{\max}=24$ the three-body width $\Gamma$ is
calculated within $30\%$ precision. Further increase of the
effective basis size is possible within the adiabatic procedure
based on the so called Feschbach reduction (FR).

\begin{figure}[ptb]
\includegraphics[width=0.47\textwidth]{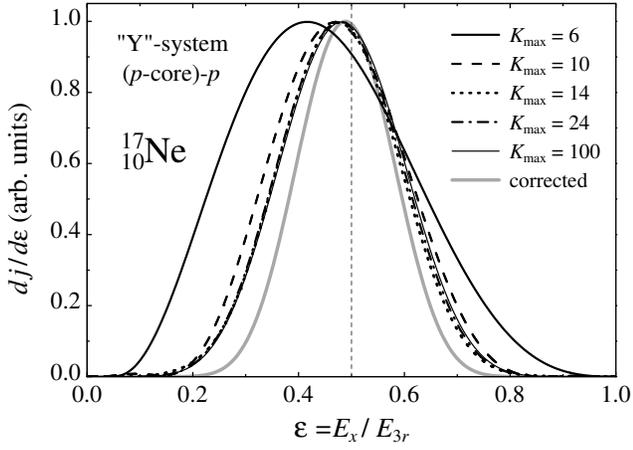}
\caption{Convergence of energy distribution for $^{17}$Ne in the ``Y'' Jacobi 
system.}
\label{fig:corcony}
\end{figure}

Feschbach reduction is a procedure, which eliminates from the total
WF $\Psi=\Psi_{p}+\Psi_{q}$ an arbitrary subspace $q$ using the
Green's function of this subspace:
\[
H_{p}=T_{p}+V_{p}+V_{pq}G_{q}V_{pq}
\]
In a certain adiabatic approximation we can assume that the radial
part of kinetic energy is small and constant under the centrifugal
barrier in the channels with so high centrifugal barrier that it is
much higher than any other interaction. In this approximation the
reduction procedure becomes trivial as it is reduced to construction
of effective three-body interactions
$V^{\text{eff}}_{K\gamma,K^{\prime}\gamma^{\prime}}$ by matrix
operations
\begin{eqnarray}
G_{K\gamma,K^{\prime}\gamma^{\prime}}^{-1} & = & (H-E)_{K\gamma,K^{\prime
}\gamma^{\prime}}=V_{K\gamma,K^{\prime}\gamma^{\prime}} \nonumber \\
 & + & \left[  E_{f}-E+\frac{(K+3/2)(K+5/2)}{2M\rho^{2}}\right]
\delta_{K\gamma,K^{\prime} \gamma^{\prime}}\,, \nonumber \\
V^{\text{eff}}_{K\gamma, K^{\prime}\gamma^{\prime}} & = &
V_{K\gamma,K'\gamma'}+\sum V_{K\gamma,\bar{K}\bar{\gamma}}
G_{\bar{K}\bar{\gamma}, \bar{K}^{\prime} \bar{\gamma}^{\prime}}
V_{\bar{K}^{\prime}\bar{\gamma}^{\prime}, K^{\prime
}\gamma^{\prime}}\;.\nonumber
\end{eqnarray}
Summation over indexes with bar is made for eliminated channels. No
strong sensitivity to the exact value of the ``Feshbach energy''
$E_{f}$ is found and we take it as $E_{f}\equiv E$ in our
calculations. More detailed account of the procedure applied within
HH method can be found in Ref.\ \cite{dan-feshb}.

It can be seen in Fig.\ \ref{fig:gam-s} (solid line) that Feschbach
reduction procedure drastically improves the convergence. However,
the calculation converges to a width value, which is somewhat
smaller than the corrected width value (that should be exact). The
reason for this effect can be understood if we make a reduction to a
smaller ``dynamic'' basis size ($K_{\max}=12$, gray line). The
calculation in this case also converges, but even to a smaller width
value. We can conclude that FR procedure allows anyhow to approach
the real width value, but provides a good result only for
sufficiently large size of the dynamic sector of the basis.

\begin{figure}[ptb]
\includegraphics[width=0.47\textwidth]{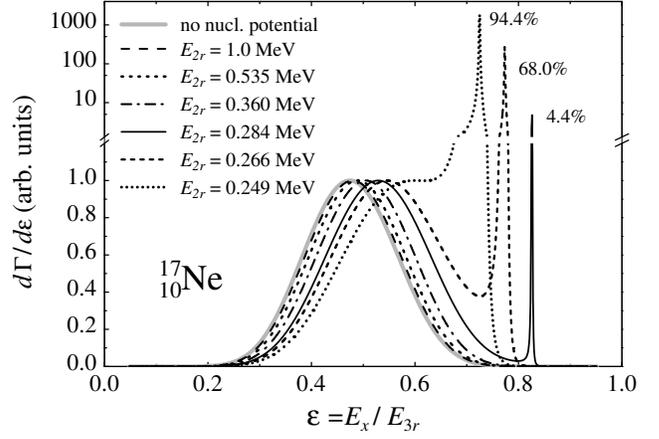}
\caption{Energy distributions for $^{17}$Ne in the ``Y'' Jacobi system for
different two-body resonance positions $E_{2r}$. The three-body decay energy is 
$E_{3r}=0.344$ MeV. The distributions
are normalized to have unity value on maximum of three-body
components. The values near the peaks show the fraction of the total
intensity concentrated within the peak. Note the change of the scale
at vertical axis.}
\label{fig:coryex}
\end{figure}

The next issue to be discussed is a convergence of the width in
calculations with different positions $E_{2r}$ of two-body resonance
in the core+$p$ subsystem. It is demonstrated for several energies
$E_{2r}$ in Fig.\ \ref{fig:gam-ex-conv}. When the resonance in the
subsystem is absent (or located relatively high) the convergence of
the width value to the exact result is very fast both in the pure
three-body and in the ``corrected'' calculation (in that case,
however, much faster). Here even FR is not required as the
convergent result is achieved in the HH calculations by
$K_{\max}=10-24$. The closer two-body resonance approaches the decay
window, the worse is convergence of HH calculations. At energy
$E_{2r}=360$ keV (which is already close to three-body decay window
$E_{3r}=344$ keV) even FR procedure provides a convergence to the
width value which is only about $65\%$ of the exact value.

In Fig.\ \ref{fig:gam-ot-ex-vx} the calculations with different
$E_{2r}$ values are summarized. The width grows rapidly as the
two-body resonance moves closer to the decay window. The
penetrability enhancement provided by the two-body resonance even
before it moves into the three-body decay window is very important.
Difference of widths with no core-$p$ FSI and FSI providing the
$s$-wave resonance to be at it experimental position $E_{2r}=0.535$
MeV is more than two orders of the magnitude. The convergence of HH
calculations also deteriorates as $E_{2r}$ moves closer to the decay
window. However, the disagreement between the HH width and the exact
value is within the order of the magnitude, until the resonance
achieves the range $E_{2r}\sim(0.7-0.85)E_{3r}$. Within this range a
transition from three-body to two-body regime happens (see also
discussion in \cite{gri03a}), which can be seen as a drastic change
of the width dependence on $E_{2r}$. This means that a sequential
decay via two-body resonance $E_{2r}$ becomes more efficient than
the three-body decay. In that case the hyperspherical expansion can
not treat the dynamics efficiently any more and the disagreement
with exact result becomes as large as orders of the magnitude. The
decay dynamics in the transition region is also discussed in details
below.

\begin{figure}[ptb]
\includegraphics[width=0.47\textwidth]{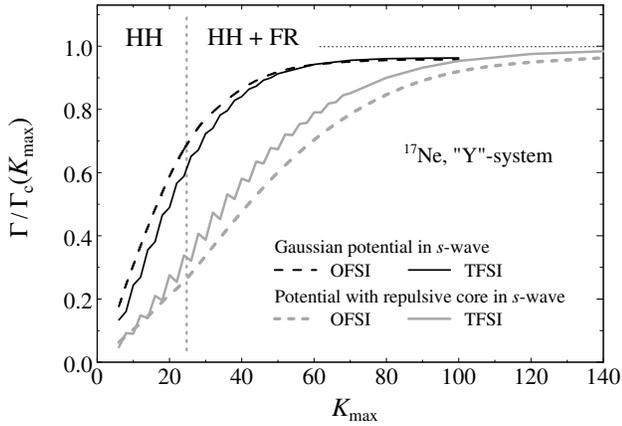}
\caption{Convergence of the $^{17}$Ne width in a simplified model. Jacobi ``Y''
system. OFSI model with $s$-wave two-body resonance at $E_{2r}=0.535$ MeV;
Gaussian potential and potential with repulsive core. TFSI model with $d$-wave
two-body resonance at $E_{2r}=0.96$ MeV.}
\label{fig:gam-nc-c}
\end{figure}

It can be seen in Fig.\ \ref{fig:gam-ot-ex-vx} that in three-body
regime the dependence of the three-body width follows well the
analytical expression
\begin{equation}
\Gamma \sim (E_{3r}/2-E_{2r})^{-2}
\label{eq:anal-e3r-e2r}
\end{equation}

The reasons of such a behaviour will be clarified in the forthcoming
paper \cite{next}. The deviations from this dependence can be found
in the decay window (close to ``transition regime'') and at higher
energies. This dependence is quite universal; e.g.\ for $^{45}$Fe it
is demonstrated in Fig.\ \ref{fig:gam-fe-ot-ex}, where it follows
the calculation results even with higher precision.

Another important issue is a convergence of energy distributions in the HH
calculations, demonstrated in Fig.\ \ref{fig:corcony} for calculations with
$E_{2r}=535$ keV. The distribution is calculated in ``Y'' Jacobi subsystem, thus
$E_{x}$ is the energy between the core and one proton. The energy distribution
convergence is fast: the distribution is stable at $K_{\max}=10-14$ and does not
change visibly with further increase of the basis. There remain a visible
disagreement with exact ("corrected") results, which give more narrow energy
distribution. We think that this effect was understood in our work
\cite{gri03c}. The three-body calculations are typically done for $\rho_{\max
}\sim500 - 2000$ fm ($\rho_{\max}\sim1000$ fm everywhere in this work). It was
demonstrated in Ref.\ \cite{gri03c} by construction of classical trajectories
that we should expect a complete stabilization of the energy distribution in
core+$p$ subsystem at $\rho_{\max}\sim 30000-50000$ fm and the effect on the
width of the energy distribution should be comparable to one observed in Fig.\
\ref{fig:corcony}.

The evolution of the energy distribution in core+$p$ subsystem with variation of
$E_{2r}$ is shown in Fig.\ \ref{fig:coryex}. When we decrease the energy
$E_{2r}$ the distribution is very stable until the two-body resonance enters the
three-body decay energy window. After that the peak at about $\varepsilon \sim
0.5$ first drifts to higher energy and then for $E_{2r} \sim 0.85 E_{3r}$ the
noticeable second narrow peak for sequential decay is formed. At $E_{2r}\sim 0.7
E_{3r}$ the sequential peak becomes so high that the three-body component of the
spectrum is practically disappeared in the background.

\begin{figure}[ptb]
\includegraphics[width=0.47\textwidth]{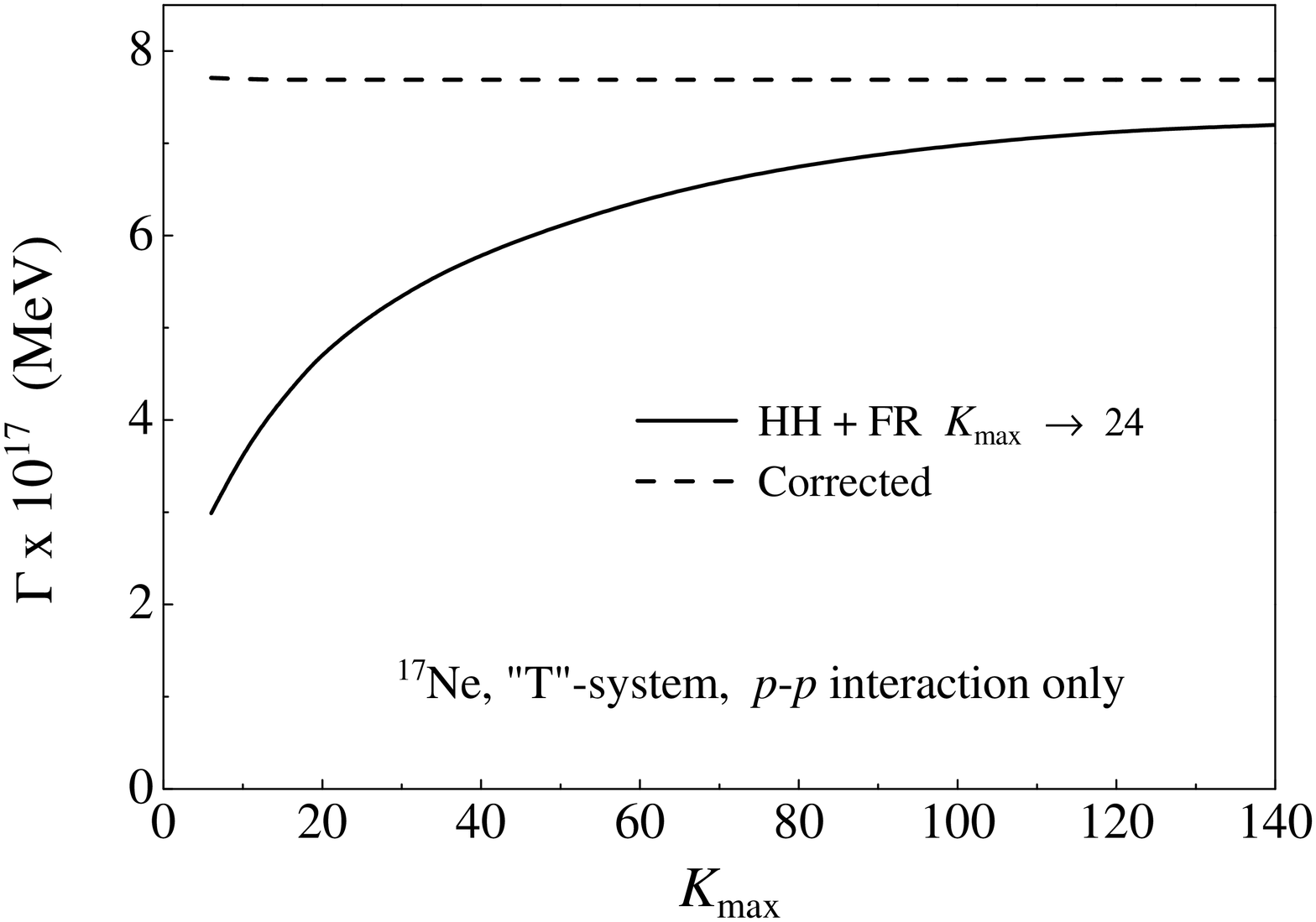}
\caption{Convergence of the $^{17}$Ne width in a simplified model.
Jacobi ``T'' system. Final state interaction describes $s$-wave
$p$-$p$ scattering.}
\label{fig:gam-all-pp}
\end{figure}

The result concerning the transition region obtained in this model is consistent
with conclusion of the paper \cite{gri03a} (where much simpler model was used
for estimates). The three-body decay is a dominating decay mode, not only when
the sequential decay is energy prohibited as $E_{2r} > E_{3r}$. Also the
three-body approach is valid when the sequential decay is formally allowed
(because $E_{2r}<E_{3r}$) but is not taking place in reality due to Coulomb
suppression at $E_{2r}\gtrsim 0.8 E_{3r}$.

Geometric characters of potentials can play an important role in the
width convergence. To test this aspect of the convergence we have
also made the calculations for potential with repulsive core. This
class of potentials was employed in studies of $^{17}$Ne and
$^{19}$Mg in Ref.\ \cite{gri03}. A comparison of the convergence of
HH calculations with $s$-wave $^{15}$O+$p$ potential from
\cite{gri03} and Gaussian potential is given in Fig.\
\ref{fig:gam-nc-c}. The width convergence in the case of the
``complicated'' potential with a repulsive core is drastically worse
than in the ``easy'' case of Gaussian potential. For typical dynamic
calculations with $K_{\max}=20-24$ the HH calculations provide only
$20-25\%$ of the width for potential with a repulsive core. On the
other hand the calculations with both potentials provide practically
the same widths $\Gamma_{c}$ \footnote{We demonstrate in paper
\cite{next} that a three-body width should depend linearly on
two-body widths of the subsystems and only very weakly on various
geometrical factors. This is confirmed very well by direct
calculations.} and FR provides practically the same and very well
converged result in both cases.


\subsection{One final state interaction --- $p$-$p$ channel}


\begin{figure}[ptb]
\includegraphics[width=0.47\textwidth]{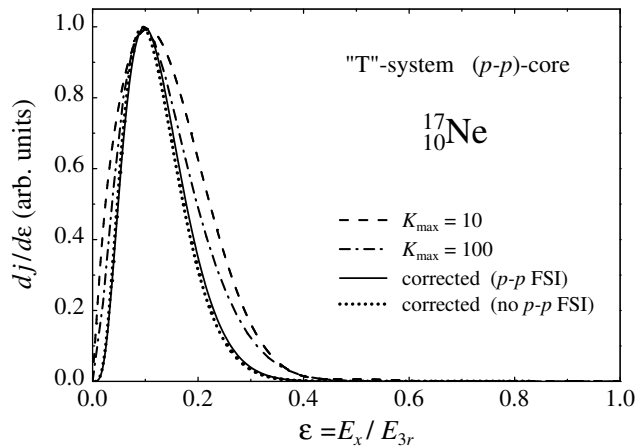}
\caption{Energy distributions for $^{17}$Ne in ``T'' Jacobi system (between two
protons).}
\label{fig:corcont}
\end{figure}

As far as two-proton decay is often interpreted as ``diproton'' decay we should
also consider this case and study how important this channel could be. For this
calculation we use a simple $s$-wave Gaussian $p$-$p$ potential, providing a
good low-energy $p$-$p$ phase shifts,
\begin{equation}
V(r)=-31\exp[-(r/1.8)^{2}]\;.
\label{pot-pp}
\end{equation}

Calculations with this potential are shown in Fig.\ \ref{fig:gam-all-pp} (see
also Table \ref{tab:wid-com-1}). First
of all the penetrability enhancement provided by $p$-$p$ FSI is much less than
the enhancement provided by core-$p$ FSI (the widths differs more than two
orders of the magnitude, see Fig.\ \ref{fig:gam-s}). This is the feature, which
has been already outlined in our works. The $p$-$p$ interaction may boost the
penetrability strongly, but only in the situation, when protons occupy
predominantly orbitals with high orbital momenta. In such a situation the
$p$-$p$ interaction allows transitions to configurations with smaller orbital
momenta in the subbarrier region, which provide a large increase of the
penetrability. In our simple model for $^{17}$Ne $3/2^{-}$ state, we have
already assumed the population of orbitals with minimal possible angular momenta
and thus no strong effect of the $p$-$p$ interaction is expected.

Also a very slow convergence of the decay width should be noted in this case.
For core-$p$ interaction the $K_{\max} \sim 10-40$ were sufficient to obtain a
reasonable result. In the case of the $p$-$p$ interaction the $K_{\max} \sim
100$ is required.

Energy distributions between two protons obtained in this model are shown in
Fig.\ \ref{fig:corcont}. Important feature of these distributions is a strong
focusing of protons at small $p$-$p$ energies. This feature is connected,
however, not with attractive $p$-$p$ FSI, but with dominating Coulomb repulsion
in the core-$p$ channel. This is demonstrated by the calculation with nuclear
FSI turned off, which provides practically the same energy distributions.
Similarly to the case of the core-$p$ FSI, very small $K_{\max}>10$ is
sufficient to provide the converged energy distribution. The converged HH
distribution is very close to the exact ("corrected") one but it is, again,
somewhat broader.

\begin{figure}[ptb]
\includegraphics[width=0.26\textwidth]{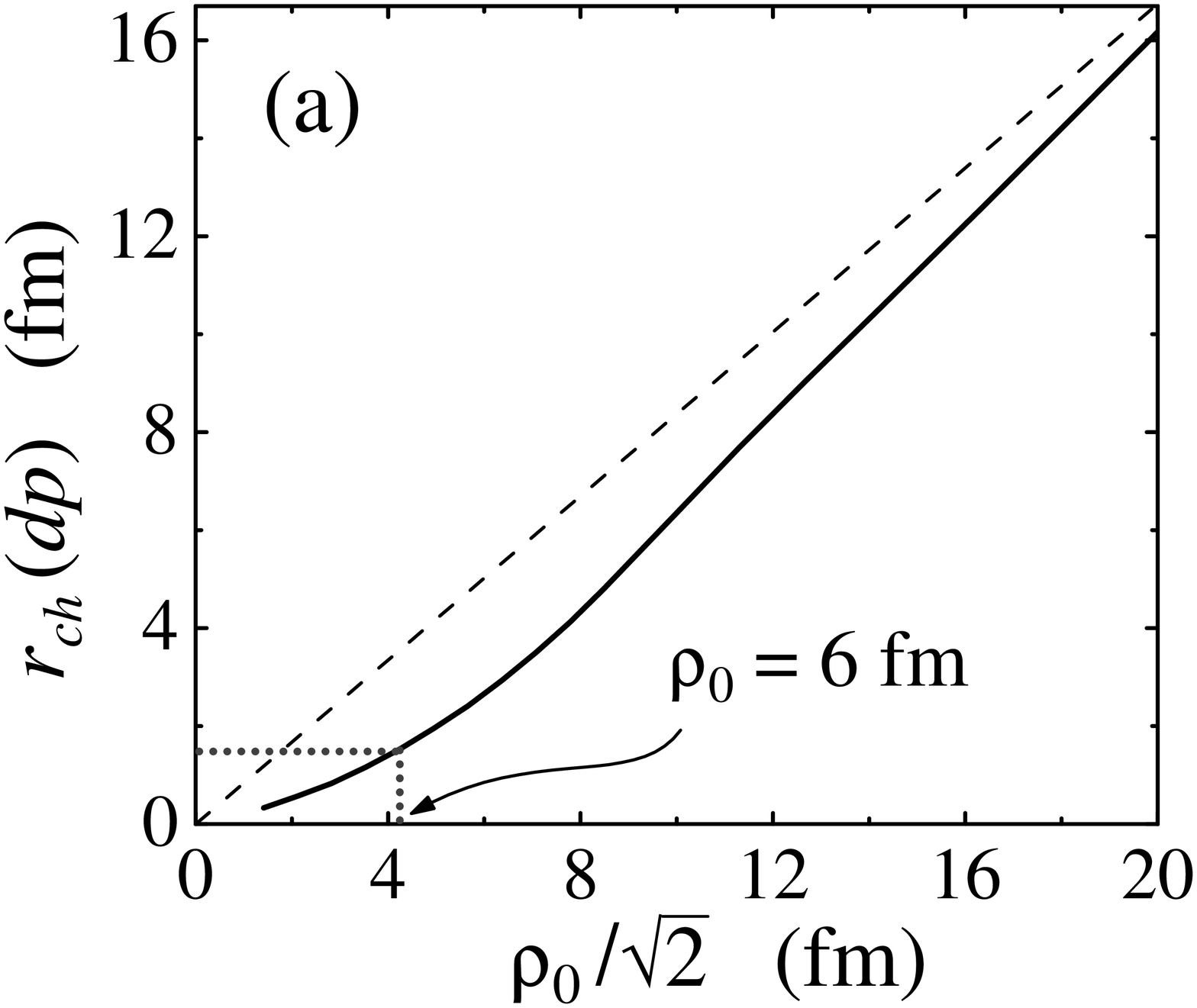}
\includegraphics[width=0.215\textwidth]{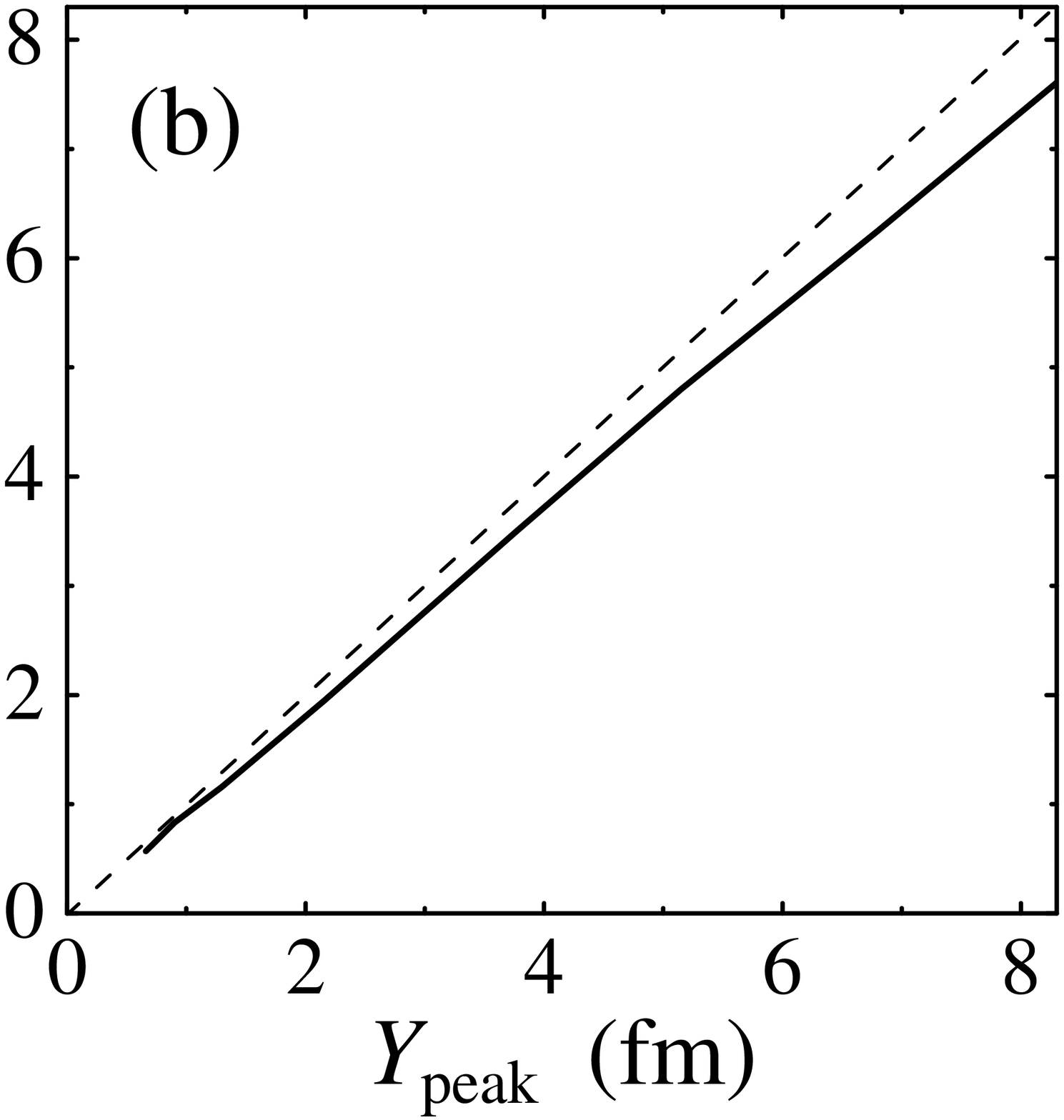}
\caption{Comparison of the OFSI calculations (solid lines) for
$^{45}$Fe in the ``T'' system with diproton model Eq.\ (\ref{dipro})
(dashed lines). Effective equivalent channel radius $r_{ch}(dp)$ for
``diproton emission'' (a) as a function of radius  $\rho_0$ of the three-body
potential (\ref{pot-3b}), the value $\rho_0/\sqrt{2}$ should
be comparable with typical nuclear sizes. (b) as a function of the
position of the peak $Y_{\text{peak}}$ in the three-body WF
$\Psi^{(+)}$ in $Y$ coordinate. The dashe lines are given to guide the eye.}
\label{fig:dipro}
\end{figure}

So far the diproton model has been treated by us as a reliable upper
limit for three-body width \cite{gri03a}. With some technical
improvements this model was used for the two-proton widths
calculations in Refs.\ \cite{bar01,bar02,bar03,bro03}. It is
important therefore to try to understand qualitatively the reason of
the small width values obtained in this form of OFSI model, which
evidently represents \emph{appropriately formulated diproton model}
\footnote{The assumed nuclear structure is very simple, but the
diproton penetration process is treated exactly --- without
assumptions about the emission of diproton from some nuclear
surface, which should be made in ``R-matrix'' approach.}. In Fig.\
\ref{fig:dipro} we compared the results of the OFSI calculations for
$^{45}$Fe in the ``T'' system with diproton width estimated by
expression
\begin{equation}
\Gamma_{dp} = \frac{1}{M_{\text{red}}r^2_{ch}(dp)} \;
P_{l=0}(0.95E_{3r},r_{ch}(dp),2Z_{\text{core}})\;,
\label{dipro}
\end{equation}
where $M_{\text{red}}$ is the reduced mass for $^{43}$Cr-$pp$ motion
and $r_{ch}(dp)$ is channel radius for diproton emission. The energy
for the relative $^{43}$Cr-$pp$ motion is taken $0.95E_{3r}$ basing
on the energy distribution in the $p$-$p$ channel (see Fig.\
\ref{fig:corcont} for example). In Fig.\ \ref{fig:dipro}a we show
the effective equivalent channel radii for diproton emission
obtained by fulfilling condition $\Gamma_{dp} \equiv \Gamma_c$ for
OFSI model calculations with different  radii $\rho_0$ of the three-body
potential Eq.\ (\ref{pot-3b}). It is easy to see that for
realistic values of these radii ($\rho_0 \sim 6 $ fm for $^{45}$Fe)
the equivalent diproton model radii should be very small ($\sim 1.5$
fm). This happens presumably because the ``diproton'' is too large
to be considered as emitted from nuclear surface of such small $\rho_0$ 
radius. Technically it can be seen as the nonlinearity of the
$r_{ch}(dp)$-$\rho_0$ dependence, with linear region achieved at
$\rho_0 \sim 15-20$ fm. Only at such unrealistically large $\rho_0$
values the typical nuclear radius (when it becomes comparable with
the ``size'' of the diproton) can be reasonably interpreted as the
surface, off which the ``diproton'' is emitted. It is interesting to
note that in the nonlinearity region for Fig.\ \ref{fig:dipro}a
there exists practically exact correspondence between the $Y$
coordinate of the WF peak in the internal region and the channel
radius for diproton emission (Fig.\ \ref{fig:dipro}b). This fact is
reasonable to interpret in such a way that the diproton is actually
emitted not from nuclear surface (as it is presumed by the existing
systematics of diproton calculations) but from the interior region,
where the WF is mostly concentrated.


\subsection{Two final state interactions}


\begin{figure}[ptb]
\includegraphics[width=0.47\textwidth]{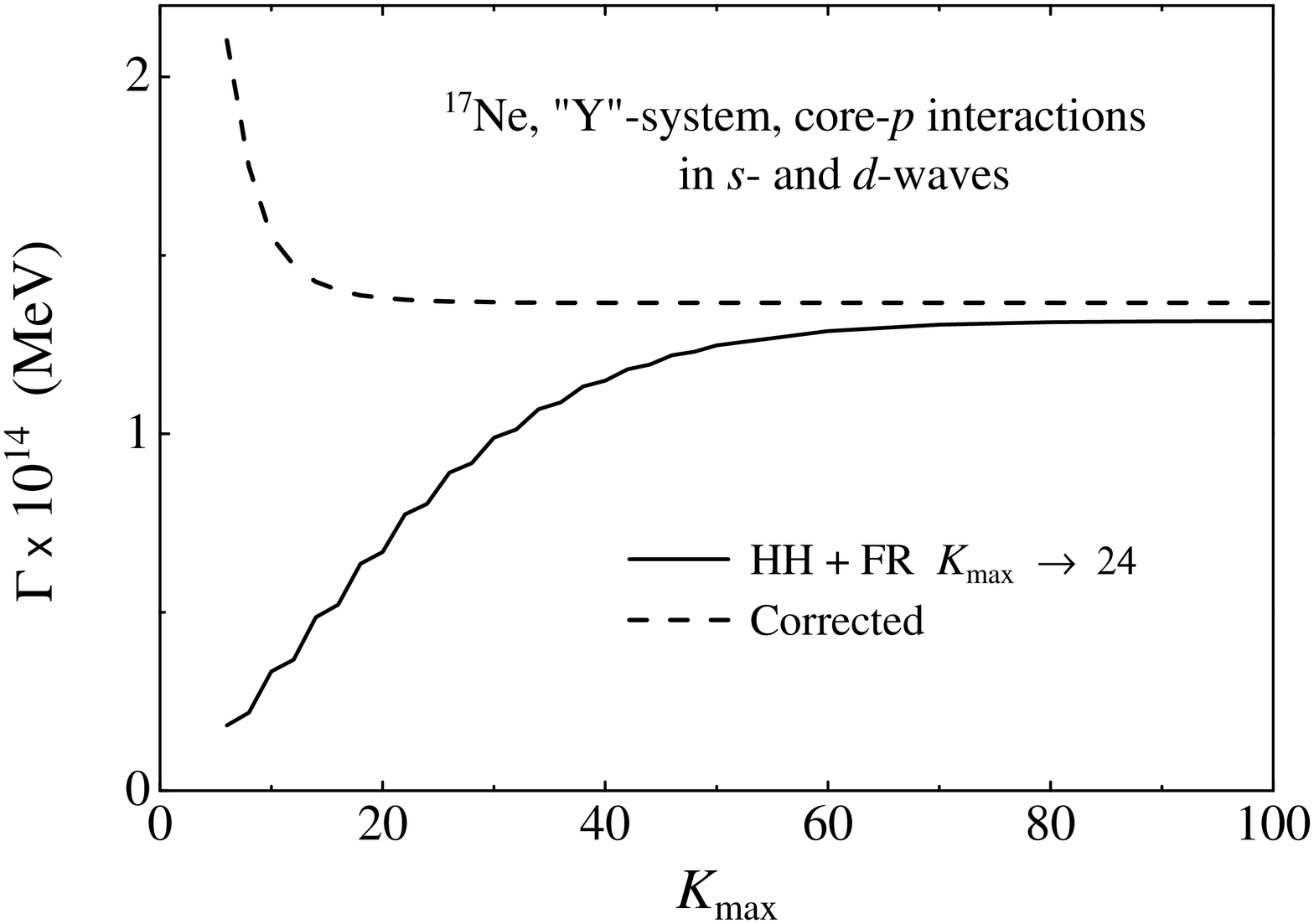}
\caption{Convergence of the $^{17}$Ne width for experimental
positions $E_{2r}=0.535$ MeV of the $0^{+}$ two-body resonance in
the ``X'' subsystem and $E_{2r}=0.96$ MeV of the $2^{+}$ two-body
resonance in the ``Y'' subsystem (TFSI model).}
\label{fig:gam-sd}
\end{figure}

As we have already mentioned the situation of one final state interaction is
comfortable for studies, but rarely realized in practice. An exception is the
case of the E1 transitions to continuum in the three-body systems, considered in
our previous work \cite{gri06}. For narrow states in typical nuclear system of
the interest there are at least two comparable final state interactions (in the
core-p channel). For systems with heavy core this situation can be treated
reasonably well as the $Y$ coordinate (in ``Y'' Jacobi system) for such systems
practically coincides with the core-$p$ coordinate. Below we treat in this way
$^{17}$Ne (for which this approximation could be not very consistent) and
$^{45}$Fe (for which this approximation should be good). In the case of
$^{17}$Ne we are thus interested in the scale of the effect, rather in the
precise width value.

For calculations with two FSI for $^{17}$Ne we used Gaussian $d$-wave potential
(see Table \ref{tab:poten-ne}), in addition to the $s$-wave potential used in
Section \ref{sec:ofsi}. This potential provides a $d$-wave state at 0.96 MeV
($\Gamma=13.5$ keV), which corresponds to the experimental position of the first
$d$-wave state in $^{16}$F. The convergence of the $^{17}$Ne decay width is
shown in Fig.\ \ref{fig:gam-sd}. Comparing with Fig.\ \ref{fig:gam-s} one can
see that the absolute value of the width has changed significantly ($2-3$ times)
but not extremely  and the convergence is practically the same. Interesting new
feature is a kind of the convergence curve ``staggering'' for odd and even
values of $K/2$. Also the convergence of the corrected calculations requires now
a considerable $K_{\max} \sim 12-14$.

\begin{figure}[ptb]
\includegraphics[width=0.47\textwidth]{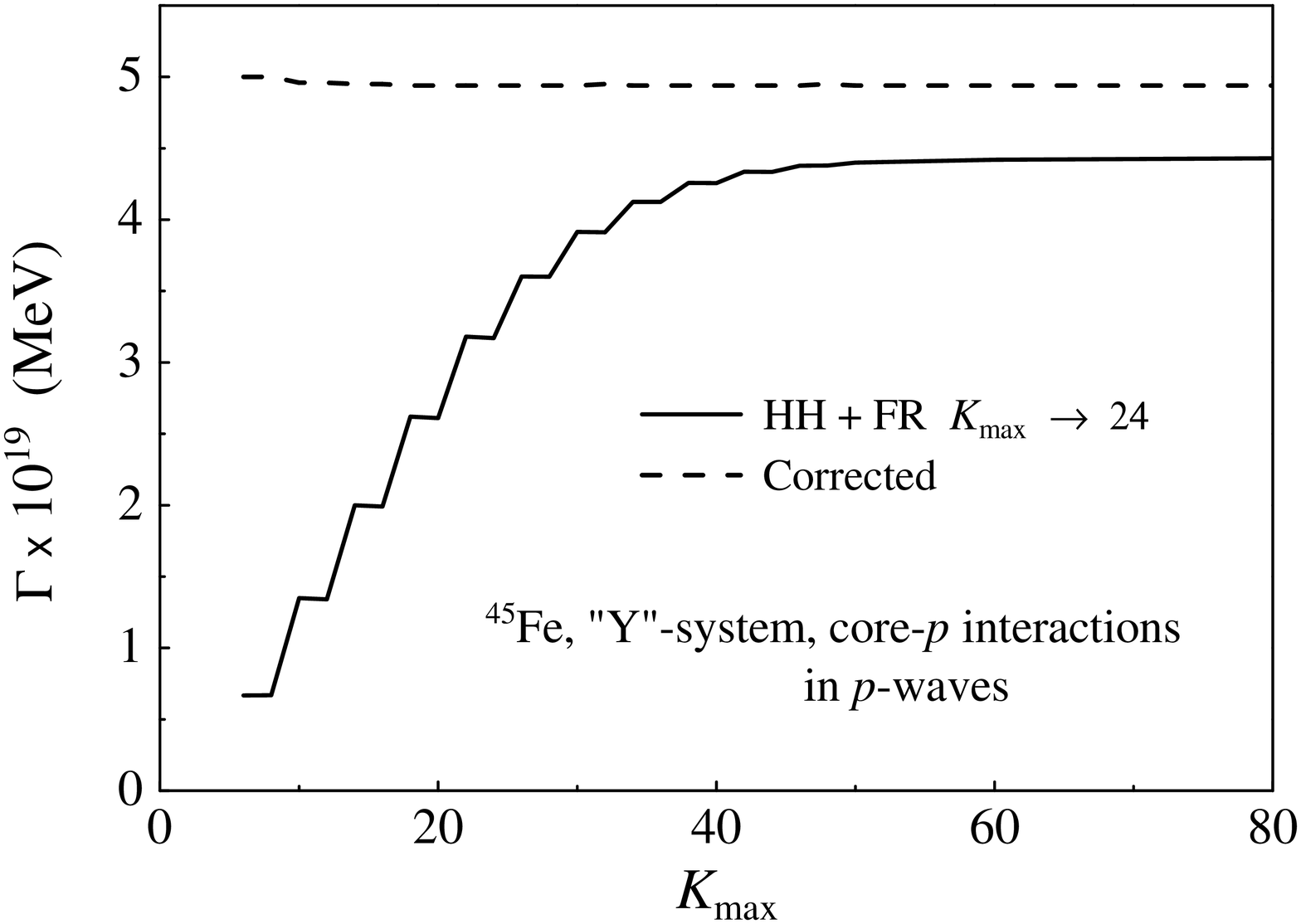}
\caption{Convergence of the $^{45}$Fe width for position of the
$1^{-}$ two-body resonances in ``X'' and ``Y'' subsystems
$E_{2r}=1.48$ MeV.}
\label{fig:gam-fe-pp}
\end{figure}

The improved experimental data for $2p$ decay of $^{45}$Fe is published recently
in Ref.\ \cite{dos05}: $E_{3r}=1.154(16)$ MeV, $\Gamma_{2p}=2.85_{-0.68}^{+0.65}
\times 10^{-19}$ MeV [$T_{1/2}(2p)=1.6_{-0.3}^{+0.5}$ ms] for two-proton
branching ratio $Br(2p)=0.57$. Below we use the resonance energy from this work.

The convergence of the $^{45}$Fe width is shown in Fig.\ \ref{fig:gam-fe-pp}.
The character of this convergence is very similar to that in the $^{17}$Ne case,
except the ``staggering'' feature is more expressed.

\begin{table}[b]
\caption{Parameters for $^{45}$Fe calculations. Potential parameters for
$p$-wave interactions (\ref{woods-saxon}) in $^{43}$Cr+$p$ channel ($V_{x0}$ in
MeV, $r_{0}=4.236$ fm, $r_{sph}=5.486$ fm) and $^{44}$Mn+$p$ ($V_{y0}$ in MeV,
$r_{0}=4.268$ fm, $r_{sph}=5.527$ fm), $a=0.65$ fm. Calculations are made with
``effective Coulomb'' of Eq.\ (\ref{app-coul-2}). Widths $\Gamma_{x}$,
$\Gamma_{y}$ of the states in the subsystems are given in keV. Corrected
three-body widths are given in the units $10^{-19}$ MeV.}
\label{tab:poten-fe}
\begin{ruledtabular}
\begin{tabular}[c]{cccccc}
$E_{2r}$ & $V_{x0}$ & $\Gamma_{x}$ & $V_{y0}$ & $\Gamma_{y}$ & $\Gamma_{c}$\\
\hline
$1.0$ & $-24.350$ & $4.3\times10^{-3}$ & $-24.54$ & $2.1\times10^{-3}$ & $26.5$
\\
$1.2$ & $-24.03$ & $0.032$ & $-24.224$ & $0.018$ & $11.8$\\
$1.48$ & $-23.58$ & $0.26$ & $-23.78$ & $0.15$ & $5.6$\\
$2.0$ & $-22.7$ & $3.6$ & $-22.93$ & $2.3$ & $2.3$\\
$3.0$ & $-20.93$ & $58$ & $-21.19$ & $44$ & $0.84$
\end{tabular}
\end{ruledtabular}
\end{table}

\begin{figure}[ptb]
\includegraphics[width=0.47\textwidth]{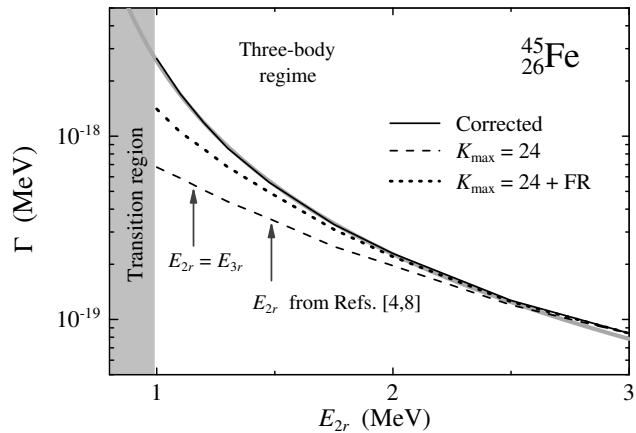}
\caption{The $^{45}$Fe g.s. width as a function of the two-body
resonance position $E_{2r}$. Dashed, dotted and solid lines show
cases of a pure HH calculation with $K_{\max}=24$, the same but with
Feshbach reduction from $K_{\max}=100$, and the corrected width
$\Gamma_{c}$. Gray area shows the transition region from three-body
to two-body decay regime. The gray curve shows simple analytical
dependence of Eq.\ (\ref{eq:anal-e3r-e2r}).}
\label{fig:gam-fe-ot-ex}
\end{figure}

The dependence of the $^{45}$Fe width on the two-body resonance energy $E_{2r}$
is shown in Fig.\ \ref{fig:gam-fe-ot-ex}. Potential parameters for these
$^{45}$Fe calculations are given in Table \ref{tab:poten-fe}. The result
calculated for $E_{3r}=1.154$ MeV and $E_{2r}=1.48$ MeV in paper \cite{gri03c}
for pure $[p^{2}]$ configuration is $\Gamma=2.85\times 10^{-19}$ MeV. The value
$K_{\max}=20$ was used in these calculations. If we take the HH width value from
Fig.\ \ref{fig:gam-fe-pp} at $K_{\max}=20$ it provides $\Gamma=2.62\times
10^{-19}$ MeV, which is in a good agreement with a full HH three-body model of
Ref.\ \cite{gri03c}. However, from Fig.\ \ref{fig:gam-fe-pp} we can conclude
that in the calculations of \cite{gri03c} the width was about $35\%$
underestimated. Thus the value of about $\Gamma=6.3\times10^{-19}$ MeV should be
expected in these calculations. On the other hand much larger uncertainty could
be inferred from Fig.\ \ref{fig:gam-fe-ot-ex} due to uncertain energy of the
$^{44}$Mn ground state. If we assume a variation $E_{2r}=1.1-1.6$ MeV the
inferred from Fig.\ \ref{fig:gam-fe-ot-ex} uncertainty of the width would be
$\Gamma=(4-16)\times 10^{-19}$ MeV. On top of that we expect a strong
$p^{2}/f^{2}$ configuration mixing which could easily reduce the width within an
order of the magnitude. Thus we can conclude that a better knowledge about
spectrum of $^{44}$Mn and a reliable structure information about $^{45}$Fe are
still required to make sufficiently precise calculations of the $^{45}$Fe width.
More detailed account of these issues is provided below.


\section{Three-body calculations}
\label{sec:three-b}

\begin{figure}
\includegraphics[width=0.47\textwidth]{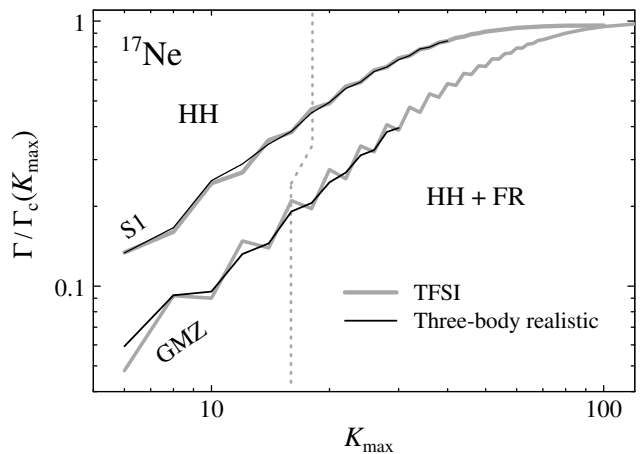}
\caption{Interpolation of $^{17}$Ne decay width obtained in full three-body
calculations by means of TFSI convergence curves (see Fig.\ \ref{fig:gam-nc-c}).
Upper curves correspond to TFSI case with Gaussian potential in $s$-wave and
compatible S1 case for full three-body model.  Lower curves correspond to TFSI
case with repulsive core potential in $s$-wave and compatible GMZ case for full
three-body model.}
\label{fig:17ne-conv-3b}
\end{figure}

Having in mind the experience of the convergence studies we have performed
large-basis calculations for $^{45}$Fe and $^{17}$Ne. They are made with
dynamical $K_{\max}=16-18$ (including Fechbach reduction from $K_{\max}=30-40$)
for $^{17}$Ne and $K_{\max}=22$ (FR from $K_{\max}=40$) for $^{45}$Fe. The
calculated width values are extrapolated using the convergence curves obtained 
in TFSI model (Figs.\ \ref{fig:17ne-conv-3b}) for $^{17}$Ne and 
\ref{fig:gam-fe-pp}for $^{45}$Fe). We have no proof that the width
convergence in the realistic three-body case is asolutely the same as in the 
TFSI case.
However, the TFSI model takes into account main dynamic features of the system
causing a slow convergence, and we are expecting that the convergence should be
nearly the same in both cases.

\begin{figure*}
\includegraphics[width=0.67\textwidth]{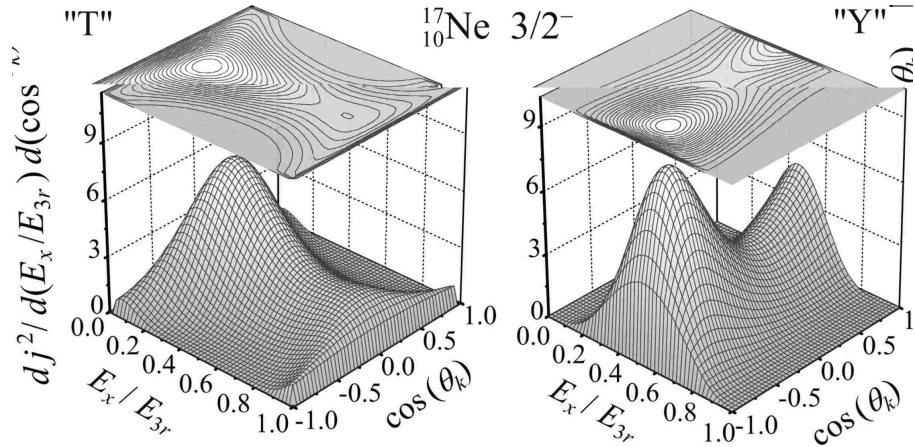}
\caption{Correlations for $^{17}$Ne decay in ``T'' and ``Y'' Jacobi systems.
Three-body calculations with realistic (GMZ) potential.}
\label{fig:corel-ne-3d}
\end{figure*}


\subsection{Widths and correlations in $^{17}$Ne}
\label{sec:three-b-ne}


The potentials used in the realistic calculations are the same as used for
$^{17}$Ne studies in Refs.\ \cite{gri03,gri05}. The GPT potential \cite{gog70}
is used in the $p$-$p$ channel. The core-$p$ potentials are referred in
\cite{gri05} as ``GMZ'' (potential introduced in \cite{gri03}) and ``high s''
(with centroid of $d$-wave states is shifted upward which is providing a higher
content of $s^2$ components in the $^{17}$Ne g.s.\ WF). Both potentials provide
correct low-lying spectrum of $^{16}$F and differ` only for $d$-wave continuum
above 3 MeV (see Table \ref{tab:16f-width}). The core-$p$ nuclear potentials,
including central, $ss$ and $ls$ terms, are taken as
\begin{eqnarray}
V(r) & = & \frac{V^{l}_{c}+({\bf s_1\cdot s_2})V^{l}_{ss}}
{1+\exp[(r-r^{l}_{0})/a]} -       ({\bf l\cdot s})\frac{ 2.0153\,
V^{l}_{ls}}{a\, r}\    \nonumber \\
& \times & \exp[(r-r^l_{0})/a] \left(1+\exp[(r-r^l_{0})/a]\right)^{-2},\;
\label{ws-core-n}
\end{eqnarray}
with parameters: $a=0.65$ fm, $r^0_0=3.014$ fm, $r^{l>0}_0=2.94$ fm,
$V^{0}_{c}=-26.381$ MeV,  $V^{1}_{c}=-9$ MeV,
$V^{2}_{c}=-57.6\;(-51.48)$ MeV, $V^{3}_{c}=-9$ MeV,
$V^{0}_{ss}=0.885$ MeV, $V^{2}_{ss}=4.5\;(12.66)$ MeV,
$V_{ls}=4.4\;(13.5)$  MeV (the values in brackets are for ``high s''
case). There are also repulsive cores for $s$- and $p$-waves
described by $a=0.4$ fm, $r^0_0=0.89$ fm, $V_{\text{core}}=200$ MeV. These
potentials are used together with Coulomb potential obtained for
Gaussian charge distribution reproducing the charge radius of
$^{15}$O.

To have extra confidence in the results, the width of the $^{17}$Ne $3/2^-$
state is calculated in several models of growing complexity (Tables
\ref{tab:wid-com}-\ref{tab:wid-com-2}). One can see from those Tables that
improvements introduced on each step provide quite smooth transition from the
very simple to the most sophisticated model.

In Table \ref{tab:wid-com} we demonstrate how the calculations in the simplified
model of Section \ref{sec:dec-simp} are compared with calculations of the full
three-body model with appropriately truncated Hamiltonian. We can switch off
corresponding interactions in the full model to make it consistent with
approximations of the simplified model. To remind, the differences of the full
model and simplified model are the following: (i) antisymmetrization between
protons is missing in the simplified model and (ii) $Y$ coordinate is only
approximately equal to the coordinate between core and second proton. Despite
these approximations the models demonstrate very close results: the worst
disagreement is not more than $30\%$.

\begin{table}[b]
\caption{Low-lying states of $^{16}$F obtained in the ``GMZ'' and ``high s''
core-$p$ potentials. The potential is diagonal in the representation with
definite total spin of core and proton $S$, which is given in the third column.}
\begin{ruledtabular}
\begin{tabular}[c]{cccccccc}
 \multicolumn{3}{c}{Case}   & \multicolumn{2}{c}{GMZ} & \multicolumn{2}{c}{high
s} & Exp.\  \\
 $J^{\pi}$ & $l$ & $S$  & $E_{2r}$ (MeV) & $\Gamma$ (keV) & $E_{2r}$ (MeV) &
$\Gamma$ (keV) & $\Gamma$ (keV) \\
\hline
$0^-$ & $0$ & $0$ & 0.535 &   18.8  &   0.535 &   18.8  &  25(5) \cite{ste06} \\
$1^-$ & $0$ & $1$ & 0.728 &   73.4  &   0.728 &   73.4  &  70(5) \cite{ste06} \\
$2^-$ & $2$ & $0$ & 0.96  &   3.5   &   0.96  &   3.5   &  6(3) \cite{ste06}  \\
$3^-$ & $2$ & $1$ & 1.2   &   9.9   &   1.2   &   10.5  &  $<15$ \cite{ajz86} \\
$2^-$ & $2$ & $1$ & 3.2   &   430   &   7.6     & $\sim 3000$   &  \\
$1^-$ & $2$ & $1$ &  4.6  &   1350  & $\sim 15$ & $\sim 6000$   &
\end{tabular}
\end{ruledtabular}
\label{tab:16f-width}
\end{table}

\begin{figure*}
\includegraphics[width=0.67\textwidth]{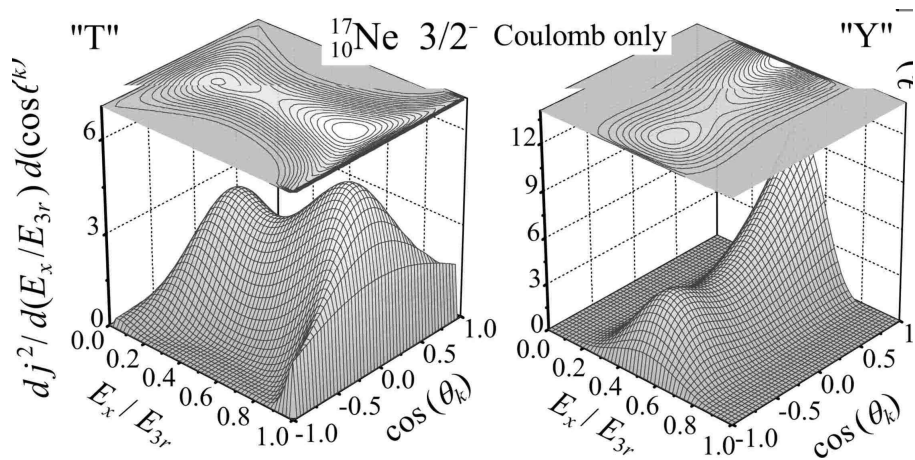}
\caption{Correlations for $^{17}$Ne decay in ``T'' and ``Y'' Jacobi systems.
Three-body calculations with Coulomb FSIs only (all nuclear pairwise potentials
are turned off).}
\label{fig:corel-nec-3d}
\end{figure*}

In Table \ref{tab:wid-com-1} we compare approximations of a
different kind: those connected with choice of the Jacobi coordinate
system in the simplified model. First we compare the ``pure
Coulomb'' case: all pairwise nuclear interactions are off and the
existence of the resonance is provided solely by the three-body
potential (\ref{pot-3b}). This model provides some hint what should
be the width of the system without nuclear pairwise interactions.
Then the models are compared with the nuclear FSIs added. The
addition of nuclear FSI drastically increase width in all cases. It
is the most ``efficient'' (in the sense of width increase) in the
case of TFSI model in the ``Y'' system. Choice of this model
provides the largest widths and can be used for the upper limit
estimates.

In Table \ref{tab:wid-com-2} full three-body models are compared.
The simplistic S1 and S2 interactions correspond to calculations
with simplified spectra of the $^{16}$F subsystem. For S1 case it
includes one $s$-wave state at 0.535 MeV ($\Gamma=18.8$ keV) and one
$d$-wave state at 0.96 MeV ($\Gamma=3.5$ keV).
These are two lower $s$- and $d$-wave states known
experimentally. In the S2 case we use instead the experimental
positions of the higher component of the $s$- and $d$-wave doublets:
$s$-wave at 0.72 MeV ($\Gamma=73.4$ keV) and $d$-wave at 1.2 MeV
($\Gamma=10$ keV). Parameters of the core-$p$ potentials can be
found in Table \ref{tab:poten-ne}. Simple Gaussian $p$-$p$ potential
(\ref{pot-pp}) is used. The variation of the results between these
models is moderate ($ \sim 30 \%$). The calculations with GMZ
potential provide the width for $^{17}$Ne $3/2^-$ state which
comfortably rests in between the results obtained in the simplified
S1 and S2 models. The structure of the WF is also obtained quite
close to these calculations. The structure in the ``high s'' case is
obtained with a strong domination of the $sd$ component. The width
in the ``high s'' case is obtained somewhat larger ($\sim 11\%$)
than in GMZ case, but this increase is consistent with the increase
of the $sd$ WF component, ($\sim 15\%$) which is expected to be more
preferable for decay than $d^2$ component.

It is important for us that the results obtained in the three-body models with
considerably varying spectra of the two-body subsystems and different
convergence
systematics appear to be quite close: $\Gamma \sim (5-8)\times 10^{-15}$ MeV.
Thus we have not found a factor which could lead to a considerable variation of
the three-body width, given the ingredients of the model are reasonably
realistic.

\begin{table}[b]
\caption{Comparison of widths for $^{17}$Ne (in $10^{-14}$ MeV
units) obtained in simplified model in ``Y'' Jacobi system and in
full three-body model with correspondingly truncated Hamiltonian.
Structure information is provided for the three-body model. In the
simplified model the weight of the $[sd]$ configuration is $100\%$
by construction. ``No $p$-$p$'' column shows the case where Coulomb
interaction in $p$-$p$ channel is switched off (see,
(\ref{app-coul-1})). ``Eff.'' column corresponds to the effective
treatment (see, (\ref{app-coul-2})) of Coulomb interaction in the
$p$-$p$ channel in the simplified model, but to the exact treatment
in full three-body model.}
\label{tab:wid-com}
\begin{ruledtabular}
\begin{tabular}[c]{ccccccc}
 & \multicolumn{2}{c}{pure Coulomb} & \multicolumn{2}{c}{OFSI} &
\multicolumn{2}{c}{TFSI} \\
 & ``no $p$-$p$'' & Eff.\ & ``no $p$-$p$'' & Eff.\  & ``no $p$-$p$'' & Eff.\ \\
\hline
Simpl.\ & 0.017 & 0.0032 & 3.02 & 0.545 & 4.70  & 1.37  \\
3-body  & 0.024\footnotemark[1] & 0.0041\footnotemark[1] & 3.22  & 0.555 & 3.91
 & 0.445  \\
$[sd]$  & 99.8  & 99.3    & 99.6  & 99.5  & 92.0  & 72.6  \\
$[p^2]$ & 0.2   & 0.6     & 0.3   & 0.4   & 0.1   & 0.2   \\
$[d^2]$ & 0     & 0       & 0     & 0     & 7.8   & 27.1  \\
\end{tabular}
\end{ruledtabular}
\footnotetext[1]{Small repulsion ($\sim 0.5$ MeV) was added in that
case in the $p$-wave core-$p$ channel to split the states with
 $sd$ and $p^2$ structure which appear practically degenerated and
strongly mixed in this model.}
\end{table}

The decomposition of the $^{17}$Ne WF obtained with GMZ potential is provided in
Table \ref{tab:struct-ne} in terms of partial internal normalizations and
partial widths. The correspondence between the components with large weights and
large partial widths is typically good. However, there are several components
giving large contribution to the width in spite of negligible presence in the
interior.

\begin{table}[b]
\caption{Comparison of widths calculated for $^{17}$Ne ($10^{-14}$ MeV units)
and $^{45}$Fe ($10^{-19}$ MeV units) with pure Coulomb FSIs and for nuclear plus
Coulomb FSIs. Simplified OFSI model in ``T'',  TFSI in ``Y'' Jacobi systems 
(``effective'' Coulomb is used in both cases) and
full three-body calculations.}
\label{tab:wid-com-1}
\begin{ruledtabular}
\begin{tabular}[c]{ccccccc}
 & \multicolumn{3}{c}{pure Coulomb} & \multicolumn{3}{c}{Nuclear+Coulomb}  \\
 &  ``T'' & ``Y'' & 3-body & ``T'' & ``Y'' & 3-body  \\
\hline
$^{17}$Ne & 0.0011 & 0.0032 & 0.0041 & 0.0077 & 1.37 & 0.76\footnotemark[1]  \\
$[sd]$    & 100    & 100    &  99.3  & 100    & 100  & 73.1  \\
$[p^2]$   & 0      & 0      &   0.6  & 0      & 0    & 1.8   \\
$[d^2]$   & 0      & 0      &    0   & 0      & 0    & 24.2   \\
$^{45}$Fe & 0.0053 & 0.0167 & 0.26 & 0.034 & 4.94 & 6.3\footnotemark[2]  \\
\end{tabular}
\end{ruledtabular}
\footnotetext[1]{This is a calculation with S1 Hamiltonian.}
\footnotetext[2]{This is a calculation providing pure $p^2$
structure.}
\end{table}

\begin{table}[b]
\caption{Width (in $10^{-14}$ MeV units) and structure of $^{17}$Ne $3/2^-$
state calculated in a full three-body model with different three-body
Hamiltonians.}
\label{tab:wid-com-2}
\begin{ruledtabular}
\begin{tabular}[c]{cccccc}
 & S1 &  S2 & GMZ & high s \\
\hline
$K_{\max}=18$         & 0.35 & 0.27 & 0.14 & 0.16      \\
Extrapolated          & 0.76 & 0.56 & 0.69 & 0.76  \\
$[sd]$  & 73.1 & 71.7 & 80.2 & 95.1 \\
$[p^2]$ & 1.8  & 1.8  & 2.0  & 1.3  \\
$[d^2]$ & 24.2 & 25.7 & 16.8 & 3.1  \\
\end{tabular}
\end{ruledtabular}
\end{table}

Complete correlation information for three-body decay of a resonant
state can be described by two variables (with omission of spin
degrees of freedom). We use the energy distribution parameter
$\varepsilon = E_x/E_{3r}$ and the angle
$\cos(\theta_k)=(\mathbf{k}_x\mathbf{k}_y)/(k_x k_y)$ between the
Jacobi momenta. The complete correlation information is provided  in
Fig.\ \ref{fig:corel-ne-3d} for realistic $^{17}$Ne $3/2^-$ decay
calculations. We can see that the profile of the energy distribution
is characterized by formation of the double-hump structure, expected
so far for $p^2$ configurations (see, e.g.\ \cite{gri03c}). This
structure can be seen both in ``T'' system (in energy distribution)
and in ``Y'' system (in angular distribution). In the calculations
of ground states of the $s$-$d$ shell nuclei we were getting such
distributions to be quite smooth. It can be found that the profile
of this distribution is defined by the $sd/d^2$ components ratio.
For example in the calculations with ``high s'' potential the total
domination of the $sd$ configuration leads to washing out of the
double-hump profile.

\begin{figure*}
\includegraphics[width=0.67\textwidth]{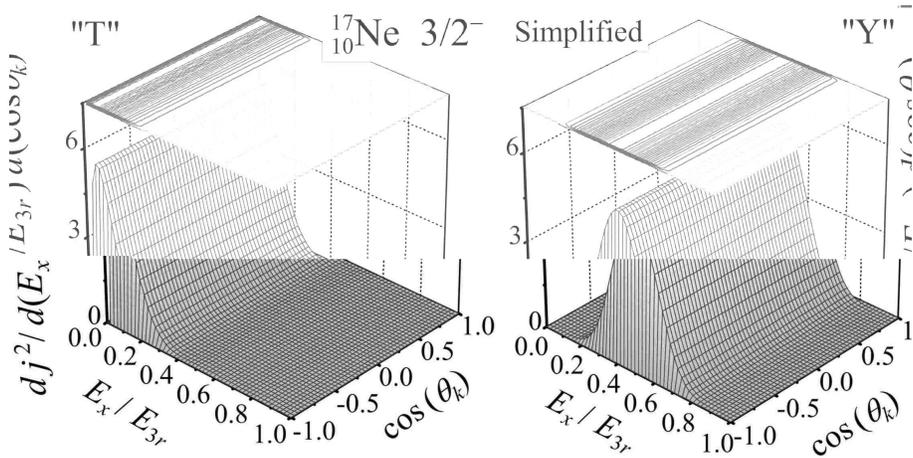}
\caption{Correlations for $^{17}$Ne decay calculated in simplified OFSI model in
``T'' (only $p$-$p$ FSI) and in ``Y'' Jacobi systems (only $s$-wave core-$p$
FSI).}
\label{fig:corel-nes-3d}
\end{figure*}

The correlations in the $^{17}$Ne (shown in Fig.\ \ref{fig:corel-ne-3d}) are
strongly influenced by the nuclear FSIs.  Calculations for only Coulomb pairwise
FSIs left in the Hamiltonan are shown in  Fig.\ \ref{fig:corel-nec-3d}. The
strong peak at small $p$-$p$ energy is largely dissolved and the most prominent
feature of the correlation density in that case is a rise of the distribution
for $\cos(\theta_k) \rightarrow 1$ in the ``Y'' Jacobi system. This kinematical
region corresponds to motion of protons in the opposite directions from the core
and is qualitatively understandable feature of the three-body Coulomb
interaction (the $p$-$p$ Coulomb interaction is minimal along such a
trajectory).

The distributions calculated in the simplified (OFSI) model are shown in Fig.\
\ref{fig:corel-nes-3d} on the same $\{\varepsilon,\cos(\theta_k) \}$ plane as in
Figs.\ \ref{fig:corel-ne-3d} and \ref{fig:corel-nec-3d}. It should be noted
that here the calculations in ``T'' and ``Y'' Jacobi systems represent different
calculations (with $p$-$p$ FSI only and with core-$p$ FSI only). In Figs.\
\ref{fig:corel-ne-3d} and \ref{fig:corel-nec-3d} two panels show different
representations of the same result. Providing reasonable (within factor $2-4$)
approximation to the full three-body model in the sense of the decay width, the
simplified model is very deficient in the sense of correlations. The only
feature of the realistic correlations which is even qualitatively correctly
described in the simplified model is the energy distribution in the ``Y''
system. The ``diproton'' model (OFSI model with $p$-$p$ interaction) fails
especially strongly, which is certainly relevant to the very small width
provided by this calculation.

\begin{table}[b]
\caption{Partial widths $\Gamma_{K\gamma}$ of different components of $^{17}$Ne
$3/2^-$ WF calculated in ``T'' Jacobi systems. Partial weights are given in
``T'' (value $N^{(T)}_{K\gamma}$) and in ``Y'' (value $N^{(Y)}_{K\gamma}$)
Jacobi systems. $S_x$ is the total spin of two protons.}
\label{tab:struct-ne}
\begin{ruledtabular}
\begin{tabular}[c]{cccccccc}
$K$ & $L$ & $l_x$ & $l_y$ & $S_x$  & $N^{(Y)}_{K\gamma}$ & $N^{(T)}_{K\gamma}$ &
$\Gamma_{K\gamma}$ \\
\hline
2  &  2 & 0 & 2 & 0 & 23.88 & 33.87  & 44.93 \\
2  &  2 & 2 & 0 & 0 & 24.97 & 16.52  & 13.29 \\
2  &  2 & 1 & 1 & 1 & 0.28  & 7.39  & 3.59  \\
2  &  2 & 1 & 1 & 0 & 1.54  &        &       \\
2  &  2 & 0 & 2 & 1 & 3.68  &        &       \\
2  &  2 & 2 & 0 & 1 & 3.68  &        &       \\
4  &  2 & 0 & 2 & 0 & 8.97  & 20.04  & 3.19  \\
4  &  2 & 2 & 0 & 0 & 8.68  & 13.57  & 5.57  \\
4  &  2 & 2 & 2 & 0 & 15.49 &  0.32  & 18.80 \\
4  &  2 & 1 & 3 & 1 & 0.03  &  2.18  & 0.95  \\
4  &  2 & 3 & 1 & 1 & 0     &  1.89  & 0.63  \\
4  &  1 & 2 & 2 & 1 & 1.02  &        &       \\
4  &  2 & 0 & 2 & 1 & 1.99  &        &       \\
4  &  2 & 2 & 0 & 1 & 2.07  &        &       \\
6  &  2 & 2 & 4 & 0 & 0.14  & 0.77   & 3.57   \\
6  &  2 & 4 & 2 & 0 & 0.14  & 0.77   & 0.78   \\
6  &  2 & 0 & 2 & 0 & 0.50  & 0.09   & 0.69    \\
8  &  2 & 4 & 4 & 0 & 0.02  & 0.003  & 1.58  \\
%
\end{tabular}
\end{ruledtabular}
\end{table}


\subsection{Width of $^{45}$Fe}
\label{sec:three-b-fe}


The calculation strategy is the same as in \cite{gri03c}. We start
with interactions in the core-$p$ channel which give a resonance in
$p$-wave at fixed energy $E_{2r}$. Such a calculation provides
$^{45}$Fe with practically pure $p^2$ structure. Then we gradually
increase the interaction in the $f$-wave, until it replaces the
$p$-wave resonance at fixed  $E_{2r}$ and then we gradually move the
$p$-wave resonance to high energy. Thus we generate a set of WFs
with different $p^2/f^2$ mixing ratios.

\begin{figure}
\includegraphics[width=0.47\textwidth]{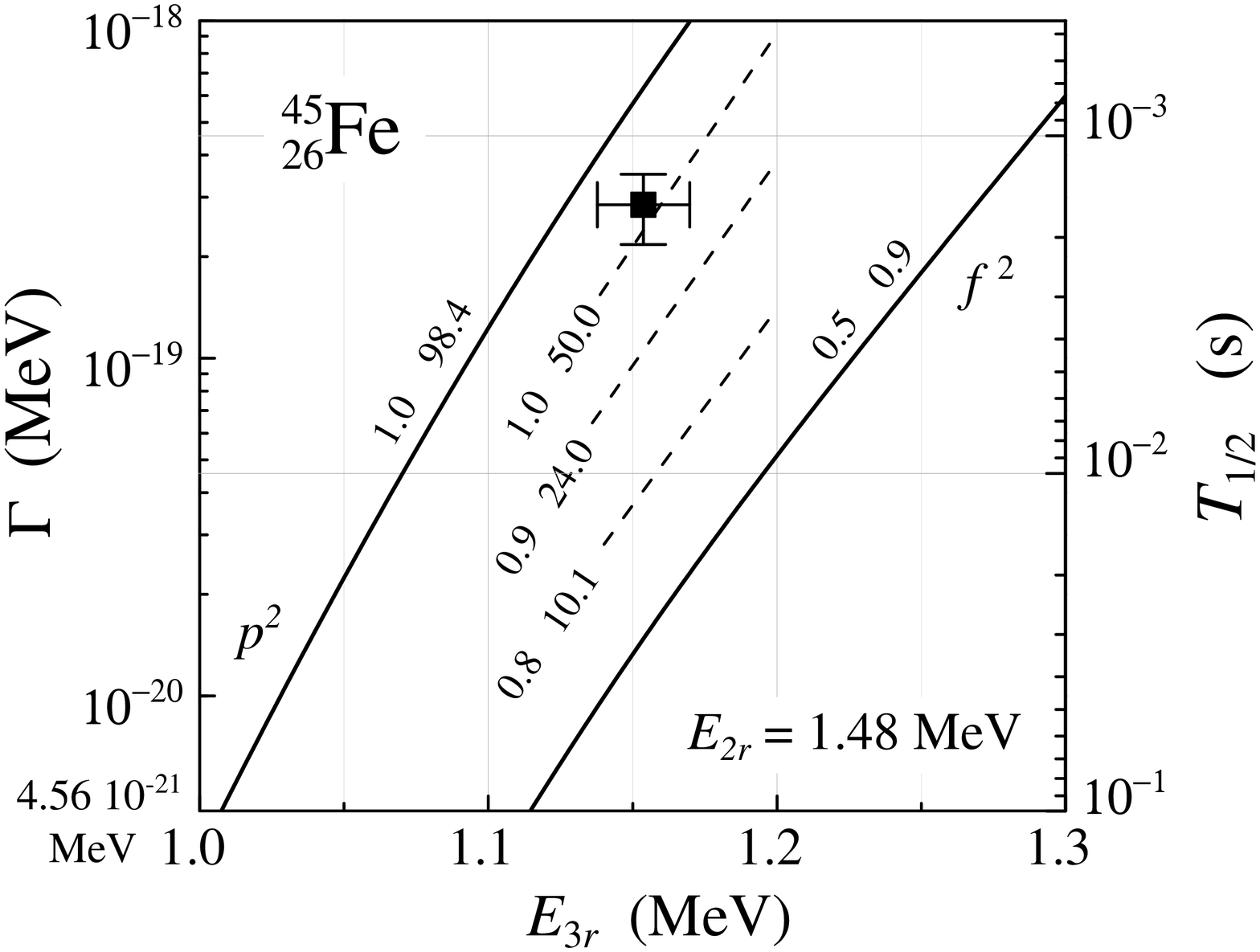}
\caption{The lifetime of $^{45}$Fe as a function of the $2p$ decay
energy $E_{3r}$. The plot is analogue of Fig.\ 6a from \cite{gri03c}
with updated experimental data \cite{dos05} and improved theoretical
results. Solid curves shows the cases of practically pure $p^2$ and
$f^2$ configurations, dashed curves stand for different mixed
$p^2/f^2$ cases.  The numerical labels on the curves show the
weights of the $s^2$ and $p^2$ configurations in percents.}
\label{fig:fe-lifetime}
\end{figure}

The results of the improved calculations with the same settings as
in \cite{gri03c} (the $^{44}$Mn g.s.\ is fixed to have $E_{2r}=1.48$
MeV) are shown in Fig.\ \ref{fig:fe-lifetime} (see also Table
\ref{tab:wid-com-1}) together with updated experimental data
\cite{dos05}. The basis size used in \cite{gri03c} was sufficient to
provide stable correlation pictures (as we have found in this work)
and they are not updated.

\begin{figure}
\includegraphics[width=0.47\textwidth]{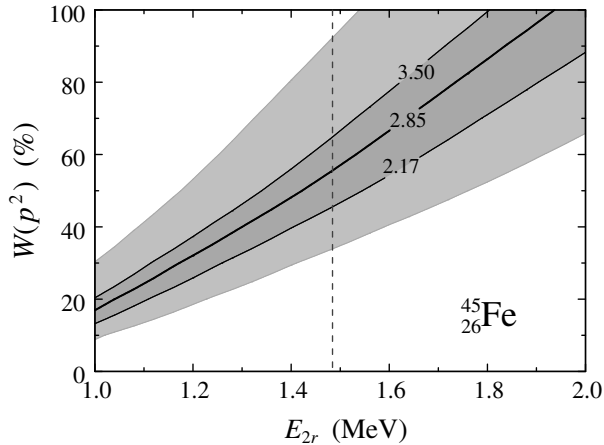}
\caption{Compatibility of the measured width of the $^{45}$Fe with different
assumptions about position $E_{2r}$ of the ground state in the $^{44}$Mn
subsystem and structure of $^{45}$Fe [weights of the $p^2$ configuration
$W(p^2)$ are shown on the vertical axis]. Central gray area corresponds to
experimental
width uncertainty $\Gamma=2.85^{+0.65}_{-0.68}\times 10^{-19}$ MeV \cite{dos05}.
The light gray area also takes into account the energy uncertainty
$E_{3r}=1.154(16)$ MeV from \cite{dos05}. The vertical dashed line corresponds 
to $E_{2r}$ used in Fig.\ \ref{fig:fe-lifetime}.}
\label{fig:wid-e2r-wp}
\end{figure}

The sensitivity of the obtained results to the experimentally
unknown energy of $^{44}$Mn can be easily studied by means of Eq.\
(\ref{eq:anal-e3r-e2r}). The results are shown in Fig.\
\ref{fig:wid-e2r-wp} in terms of the regions consistent with
experimental data on the $\{E_{2r},W(p^2)\}$ plane [$W(p^2)$ is the
weight of $p^2$ configuration in $^{45}$Fe WF]. It is evident from
this plot that our current experimental knowledge is not sufficient
to draw definite conclusions. However, it is also clear that with
increased precision of the lifetime and energy measurements for
$^{45}$Fe and the appearance of more detailed information on
$^{44}$Mn subsystem the restrictions on the theoretical models
should become strong enough to provide the important structure
information.


\section{Discussion}
\label{sec:disc-1}


General trends of the model calculations can be well understood from
Tables \ref{tab:wid-com}-\ref{tab:wid-com-2}. For the pure Coulomb
case the simplified model calculations (in the ``Y'' and ``T''
systems) and three-body calculations provide reasonably consistent
results. The simplified calculations in the ``Y'' system always give
larger widths than those in the ``T'' system. From decay dynamics
point of view this leads to understanding of the contradictory fact
that the sequential decay path is preferable even if no even virtual
sequential decay is possible (as the nuclear interactions are
totally absent in this case).

The calculations with attractive nuclear FSIs rather expectedly
provide larger widths than the corresponding calculations with
Coulomb interaction only. The core-proton FSI is much more efficient
for width enhancement than $p$-$p$ FSI. This fact is correlated with
the observation of the previous point and is a very simple and
strong indication that the wide-spread perception of the two-proton
decay as ``diproton'' decay is to some extent misleading. As it has
already been mentioned the $p$-$p$ FSI influences the penetration
strongly in the very special case when the decay occurs from
high-$l$ orbitals (e.g.\ $f^2$ in the case of $^{45}$Fe). Thus we
should consider as not fully consistent the attempts to explain
two-proton decay results only by the FSI in the $p$-$p$ channel
(e.g.\ Ref.\ \cite{bro03}) as much stronger decay mechanism is
neglected in these studies.

From techical point of view the states considered in this work
belong to the most complicated cases. The complication is due to the
ratio between the decay energy and the strength of the Coulomb
interaction (it defines the subbarrier penetration range to be
considered dynamically). Thus the convergence effects demonstrated
in this work for $^{17}$Ne have the strongest character among the
systems studied in our previous works
\cite{gri02a,gri03,gri03a,gri03c}. Because of the relatively small
$K_{\max}=12$ used in the previous works we have found an order of
the magnitude underestimation of the $^{17}$Ne($3/2^{-}$) width. For
systems like $^{48}$Ni --- $^{66}$Kr the underestimation of widths
in our previous calculations is expected to be about factor of 2. A
much smaller effect is expected for lighter systems.

It was demonstrated in \cite{gri05a,gri06} that the capture rate for
the $^{15}$O($2p$,$\gamma$)$^{17}$Ne reaction depends strongly on
the two-proton width of the first excited $3/2^{-}$ state in
$^{17}$Ne. This width was calculated in Ref.\ \cite{gri03} as $4.1
\times10^{-16}$ MeV (some confusion can be connected with misprint
in Table III of Ref.\ \cite{gri03}, see erratum). However, in the
subsequent work \cite{gar04}, providing very similar to \cite{gri03}
properties of the $^{17}$Ne WFs for the ground and the lowest
excited states, the width of the $3/2^{-}$ state was found to be
$3.6 \times10^{-12}$ MeV. It was supposed in \cite{gar04} that such
a strong disagreement is connected with poor subbarrier convergence
of the HH method in \cite{gri03} compared to Adiabatic Faddeev HH
method of \cite{gar04}. This point was further reiterated in Ref.\
\cite{gar05}. We can see now that this statement has a certain
ground. However, the convergence problems of the HH method are far
insufficient to explain the huge disagreement: the width increase
found in this work is only one order of magnitude. The most
conservative upper limit $\Gamma \sim 5 \times 10^{-14}$ MeV (see
Table \ref{tab:wid-com}) was obtained in a TFSI calculation
neglecting $p$-$p$ Coulomb interaction. The other models
systematically produce smaller values, with realistic calculations
confined to the narrow range $\Gamma \sim (5-8) \times10^{-15}$ MeV
(Table \ref{tab:wid-com-2}). Thus the value $\Gamma \sim 4
\times10^{-12}$ MeV obtained in paper \cite{gar04} is very likely to
be erroneous. That result is possibly connected
 with a simplistic quasiclassical procedure for width calculations employed
in this work.


\section{Conclusion.}


In this work we derive the integral formula for the widths of the
resonances decaying into the three-body channel for simplified
Hamiltonians and discuss various aspects of its practical
application. The basic idea of the derivation is not new, but for
our specific purpose (precision solution of the multichannel
problem) several important features of the scheme have not been
discussed.

We can draw the following conclusions from our studies.

\noindent(i) We presume that HH convergence in realistic
calculations should be largely the same as in the simplified
calculations as they imitate the most important dynamic aspects of
the realistic situation. The width values were somewhat
underestimated in our previous calculations. The typical
underestimation ranges from few percent to tens of percent for
``simple'' potential and from tens of percent to an order of
magnitude in ``complicated'' cases (potentials with repulsive core).

\noindent(ii) Convergence of the width calculations in the
three-body HH model can be drastically improved by a simple
adiabatic version of the Feshbach reduction procedure. For a
sufficiently large dynamic sector of the basis the calculation with
effective FR potential converges from below and practically up to
the exact value of the width. For a small dynamic basis the FR
calculation converges towards a width value smaller than the exact
value, but still improves considerably the result.

\noindent(iii) The energy distributions obtained in the HH
calculations are quite close to the exact ones. Convergence with
respect to basis size is achieved at relatively small $K_{\max}$
values. The disagreement with exact distributions is not very
significant and is likely to be connected not with basis size
convergence but, with radial extent of the calculations
\cite{gri03c}.

\noindent(iv) Contributions of different decay mechanisms were
evaluated in the simplified models. We have found that the
``diproton'' decay path is much less efficient than the
``sequential'' decay path. This is true even in the model
calculations without nuclear FSIs (no specific dynamics), which
means that the ``sequential'' decay path is somehow kinematically
preferable.

\noindent(v) The value of the width for $^{17}$Ne $3/2^{-}$ state
was underestimated in our previous works by around an order of
magnitude. A very conservative upper limit is obtained in this work
as $\Gamma\sim 5 \times10^{-14}$ MeV, while typical values for
realistic calculations are within the  $(5-8)\times 10^{-15}$ MeV
range. Thus the value $\Gamma\sim 4 \times10^{-12}$ MeV obtained in
papers \cite{gar04,gar05} is likely to be erroneous.

From this paper it is clear that the convergence issue is
sufficiently serious, and in some cases were underestimated in our
previous works. However, from practical point of view, the
convergence issue is not a principle problem. For example the
uncertain structure issues and subsystem properties impose typically
much larger uncertainties for width values. For heavy two-proton
emitters (e.g.\ $^{45}$Fe) the positions of resonances in the
subsystems are experimentally quite uncertain. For a moment this is
the issue most limiting the precision of theoretical predictions. We
have demonstrated that with increased precision the experimental
data impose strong restrictions on theoretical calculations allowing
to extract an important structure information.


\section{Acknowledgements}


The authors are grateful to Prof.\ K. Langanke and Prof.\ M.\ Ploszajczak for
interesting discussions. The authors acknowledge the financial support from the
Royal Swedish Academy of Science. LVG is supported INTAS Grants 03-51-4496 and
05-1000008-8272, Russian RFBR Grants Nos.\ 05-02-16404 and 05-02-17535 and
Russian Ministry of Industry and Science grant NS-8756.2006.2.

%

%

%

%




\end{document}